# Steps towards nonlinear cluster inversion through gravitational distortions.
# I. Basic considerations and circular clusters

Peter Schneider & Carolin Seitz

## Abstract:

The distortion of images of faint, high-redshift galaxies by light deflection at foreground clusters of galaxies can be used to determine the (projected) mass distribution of the clusters. In the case of strong distortions, which lead to arcs in clusters, the position of the arc and/or its radius of curvature yields an estimate for the total mass inside a circle traced out by the arc. Weak distortions, which can be observed to much larger angular separations, can be used to determine the mass profile. In the case of weak distortions, an approximation which identifies the observed distortion with the shear produced by the lens can be made; this (linear) approximation breaks down, however, if one wants to probe the center of the cluster, i.e., approach the region in which giant arcs can be formed. The methods devolped hitherto, the most advanced of which is that by Kaiser & Squires, rely on the linear approximation and can thus yield no reliable results for regions close to the center of the cluster. The purpose of this paper is to provide a theoretical basis to generalize the Kaiser & Squires method to the nonlinear regime, thereby making it more powerful, since the nonlinear distortions provide strong constraints on the mass profile, which in combination with the weak distortion should yield more reliable cluster inversions.

We discuss the statistical properties of the observable image ellipticities and provide several methods to determine the local distortion of the lens from observed galaxy images, some of which rely only on the *orientation* of the images, not on their ellipticity. Analytic approximations for the error in the locally determined distortion are provided. An invariance transformation of the density profile of the cluster is derived which leaves the observable distortion invariant; this transformation differs from that valid in the linear approximation where it is simply the addition of a homogeneous matter sheet. We then investigate the inversion of spherical clusters; in particular, we show the precision with which the center of a spherical cluster can be determined, both with and without usage of the absolute value of the image ellipticities. We show that inclusion of the absolute value of the ellipticities increases the accuracy of the center position considerably. Finally, we study the fully nonlinear inversion of spherical clusters in some detail, to illustrate the difficulties one has to anticipate in nonlinear reconstructions of realistic two-dimensional matter distributions. We compare the direct nonlinear inversion with the method of fitting parametrized mass profiles to the distortion data, and point out relative strengths and weaknesses. In particular, we emphasise the influence of the above-mentioned global invariance transformation and an additional local degeneracy.

# 1 Introduction

Since the first discovery of lensing in clusters (Lynds & Petrosian 1986, Soucail et al. 1987) and the interpretation of giant luminous arcs as gravitationally distorted images of background galaxies (Paczynski 1987), which was later verified by the redshift measurements of arcs (e.g., Soucail et al. 1988), it was realized that these lensing events provide a unique means to study the mass distribution in clusters. Although a single luminous arc can only determine the mass enclosed in a circle traced out by the arc, the observation of several arclets (Fort et al. 1988) in the cluster A370 can be used to set more constraints on the mass distribution of this cluster (Grossman & Narayan 1989, Kochanek 1990). The pioneering observations by Tyson, Valdes & Wenk (1990) demonstrated that it is feasible to measure even weaker distortions of distant galaxies caused by a foreground cluster, and that these weak distortions can be used to trace the surface mass distribution of the cluster. A detailed study of these possibilities has been performed by Kochanek (1990) and Miralda-Escude (1991, ME91 hereafter), where the main emphasis was put on determining the profile of spherical clusters. Kneib et al. (1994) have tried to determine the redshift of several arclets in the cluster A370, for which a fairly detailed lens models has been constructed by Kneib et al. (1993), by calculating that redshift of the source which minimizes the intrinsic ellipticity of the source, given its observed image ellipticity and location in the cluster. For a recent review on cluster lenses, see Fort & Mellier (1994).

Major progress then came through the work of Kaiser & Squires (1993, hereafter KS), in which an inversion formula for the mass profile of a cluster was derived and tested on numerical data and the observations of Tyson et al. (1990). Weak distortions have been verified around the double QSO 2345+007 (Bonnet et al. 1993), increasing the evidence for this system to be a gravitational lens system. The application of the Kaiser & Squires method to several clusters (Fahlman et al. 1994, Smail et al. 1994a,b) have shown the principle potential of this method, but has also disclosed some of its weaknesses. For example, the direct reconstruction method as proposed in KS yields fairly noisy matter distributions.

This is the first paper of a series in which we intent to study the problem of cluster inversion in more detail. In particular, the method of Kaiser & Squires applies specifically to the case of weak lensing, and thus the constraints from stronger distortions (i.e., arcs and arclets) cannot be easily merged with their inversion technique. More specifically, the method of KS is based on the fact that the shear distribution caused by a deflector depends linearly on the surface mass density; this linear relation can be inverted. However, the local shear cannot be obtained from image distortions; only in the case of weak lensing can the shear be identified with the local distortion. In the more general case which includes strong distortions, the relation between shear and distortion is nonlinear. One of the goals of this work is to find an inversion method which allows the inclusion of nonlinear effects. Furthermore, whereas the KS inversion equation is exact in the linear regime, its application to an observed sample of distorted images is burdened by several problems: observation of a finite number of galaxies introduces a shot noise to the inverted density profile. The prize for dealing with that noise is the finite resolution produced by a smoothing process. As we shall demonstrate in our next paper (Seitz & Schneider, in preparation), the smoothing procedure proposed by KS can be significantly improved. Second, the inversion formula requires data over the whole lens plane, whereas real observations are limited by the size of the CCD. In applying the inversion formula, an extrapolation of the distortion field into the unobserved part of the sky must be performed; equivalently, the surface mass density at the edge of the observed field must be specified. In either of these precedures, edge effects occur, which are most clearly visible in Fig. 7 of Smail et al. (1994b). These edge effects can be reduced by choosing a 'reasonable' extrapolation of the distortion field.

This first paper is confined to some basic considerations of the distortion of images by a foreground gravitational lens. In order to concentrate on the essentials, we will make the approximation that all sources (galaxies) have the same redshift, or that the distance ratio $D_{\rm ds}/D_{\rm s}$ (see Sect. 2) is the same for all sources. This assumption is not valid in reality, though it may be a fair approximation for lensing clusters at relatively low redshift. In Sect. 2 we provide the basic equations for the transformation of source ellipticities and image ellipticities and introduce our notation. Section 3 deals with the local observables, which mainly is the distortion, but not the shear. In fact, we shall see that even if the local surface mass density were known, the shear cannot be uniquely determined locally from the local distortion, but a degeneracy occurs. This degeneracy provides one of the main obstacles for the nonlinear cluster inversion. The local distortion can be determined from a sample of images; we present a method how this can be done in principle, and derive a relation between the local distortion and the mean image ellipticity, as well as its dispersion. Fit formulae for both quantities are presented which simplify their application. We also consider the statistical properties of the image alignments, and show that one can determine the local distortion also from the alignments only, without referring to the absolute values of the image ellipticities. This point deserves particular attention, since the observational determination of the image alignments is much less affected by observing conditions (most notably seeing) than the absolute value of the ellipticity. We then derive an invariance transformation of the surface mass density which leaves the distribution of the distortions invariant; this invariance transformation generalizes that obtained by Kaiser & Squires, valid in the linear regime, which is simply the addition of a homogeneous sheet of matter. In fact, the invariance transformation derived here corresponds to the 'magnification transformation' of Gorenstein, Falco & Shapiro (1988).

The next two sections deal with an application to spherical clusters. In Sect. 4 we consider several methods for determining the center of a spherical cluster from image distortions and apply them to synthetic data. It is demonstrated that for critical clusters, the center can be determined much more accurately than for noncritical ones, if one makes use also of more strongly distorted images. In Sect. 5 we consider three fairly different methods for obtaining the density profile of a spherical cluster. We compare these methods, one of which is similar to the method of Kaiser & Squires specialized to a spherical cluster, whereas the other two are based on parametrized models, and discuss their relative advantages and potential drawbacks. Our results are briefly summarized in Sect. 6.

## 2 Imaging of extended sources

We can describe an extended source by its matrix of second moments: let $I_{\rm s}(\boldsymbol{\theta})$ be the surface brightness distribution of the source, and let $q(I)$ be a function of the surface brightness, we define the center of the source by (see Blandford et al. 1991)

$$\bar{\boldsymbol{\beta}} = \frac{\int \mathrm{d}^2\beta \, q(I_\mathrm{s}(\boldsymbol{\beta})) \, \boldsymbol{\beta}}{\int \mathrm{d}^2\beta \, q(I_\mathrm{s}(\boldsymbol{\beta}))} \quad , \tag{2.1}$$

and quadrupole matrix of the source,

$$Q_{ij}^{(\mathrm{s})} = \frac{\int \mathrm{d}^2\beta \, q(I_\mathrm{s}(\boldsymbol{\beta})) \, (\beta_i - \bar{\beta}_i)(\beta_j - \bar{\beta}_j)}{\int \mathrm{d}^2\beta \, q(I_\mathrm{s}(\boldsymbol{\beta}))} \quad . \tag{2.2}$$

The function $q(I)$ can be chosen appropriately; for example, it can be a Heaviside step function, in which case the shape of the source within a limiting isophote contour would be obtained; alternatively, $q(I)$ can be chosen to give more weight to the center of the source which, however, can be strongly affected by seeing effects, or to the outer regions of a source which is probably less sensitive to seeing effects, but likely to be more noisy. The choice of $q$ has to be tested on real data. As an example, we consider a source with elliptical isophote contours, centered on $\boldsymbol{\beta} = 0$,

$$I_\mathrm{s}(\boldsymbol{\beta}) = J\left(\frac{\beta_1'^2}{a^2} + \frac{\beta_2'^2}{b^2}\right) \quad ,$$

with $b/a \leq 1$ being the axis ratio, $\boldsymbol{\beta} = \mathcal{R}(\vartheta)\boldsymbol{\beta}'$,

$$\mathcal{R}(\vartheta) = \begin{pmatrix} \cos\vartheta & -\sin\vartheta \\ \sin\vartheta & \cos\vartheta \end{pmatrix} \tag{2.3}$$

describing the orientation of the image, such that the major axis encloses an angle $\vartheta$ with the positive $\beta_1$-axis, and $J$ being an arbitrary function describing the profile of the source. Then,

$$Q_{11}^{(\mathrm{s})} = Q_0 \left(a^2 \cos^2\vartheta + b^2 \sin^2\vartheta\right) \quad ,$$
$$Q_{12}^{(\mathrm{s})} = Q_{21}^{(\mathrm{s})} = Q_0 \left(a^2 - b^2\right) \cos\vartheta \, \sin\vartheta \quad ,$$
$$Q_{22}^{(\mathrm{s})} = Q_0 \left(a^2 \sin^2\vartheta + b^2 \cos^2\vartheta\right) \quad ,$$

and

$$Q_0 = \frac{\int \mathrm{d}t \, t^3 \, q\left(J\left(t^2\right)\right)}{2 \int \mathrm{d}t \, t \, q\left(J\left(t^2\right)\right)}$$

describes the 'size' of the source.

To describe the shape (including orientation) of a source, we can define a complex number $\chi^{(\mathrm{s})}$, given by

$$\chi^{(\mathrm{s})} = \frac{\left(Q_{11}^{(\mathrm{s})} - Q_{22}^{(\mathrm{s})}\right) + 2\mathrm{i}Q_{12}^{(\mathrm{s})}}{Q_{11}^{(\mathrm{s})} + Q_{22}^{(\mathrm{s})}} \quad . \tag{2.4a}$$

The *ellipticity* $\chi^{(\mathrm{s})}$ is given in terms of the eigenvalues $\lambda_1 \geq \lambda_2 \geq 0$ of $Q^{(\mathrm{s})}$ as

$$\chi^{(\mathrm{s})} = \frac{\lambda_1 - \lambda_2}{\lambda_1 + \lambda_2} \mathrm{e}^{2\mathrm{i}\vartheta_\mathrm{s}} \quad , \tag{2.4b}$$

where $\vartheta_\mathrm{s}$ is the position angle of the eigenvector belonging to the larger eigenvalue $\lambda_1$. For a source with elliptical isophotes,

$$\chi^{(\mathrm{s})} = \frac{1-(b/a)^2}{1+(b/a)^2}\,\mathrm{e}^{2\mathrm{i}\vartheta_{\mathrm{s}}} \quad .$$

If a sufficiently small source is imaged by a gravitational lens, it can be described by the linearized lens mapping

$$\mathrm{d}\boldsymbol{\beta} = A(\boldsymbol{\theta})\,\mathrm{d}\boldsymbol{\theta} \quad , \tag{2.5}$$

where

$$A(\boldsymbol{\theta}) = \begin{pmatrix} 1-\kappa-\gamma_1 & -\gamma_2 \\ -\gamma_2 & 1-\kappa+\gamma_1 \end{pmatrix} \tag{2.6}$$

is the Jacobian matrix of the lens equation

$$\boldsymbol{\beta} = \boldsymbol{\theta} - \boldsymbol{\alpha}(\boldsymbol{\theta}) \quad . \tag{2.7}$$

Here, $\boldsymbol{\alpha}(\boldsymbol{\theta})$ is the (scaled) deflection angle of a light ray at position $\boldsymbol{\theta}$, given in terms of the surface mass density $\Sigma(D_\mathrm{d}\boldsymbol{\theta})$ of the lens as

$$\boldsymbol{\alpha}(\boldsymbol{\theta}) = \frac{1}{\pi}\int_{\mathrm{I\!R}^2}\mathrm{d}^2\theta'\,\kappa(\boldsymbol{\theta}')\,\frac{(\boldsymbol{\theta}-\boldsymbol{\theta}')}{|\boldsymbol{\theta}-\boldsymbol{\theta}'|^2} \quad , \tag{2.8}$$

with the dimensionless surface mass density

$$\kappa(\boldsymbol{\theta}) = \frac{\Sigma(D_\mathrm{d}\boldsymbol{\theta})}{\Sigma_\mathrm{crit}} \tag{2.9a}$$

and the critical surface mass density

$$\Sigma_\mathrm{crit} = \frac{c^2 D_\mathrm{s}}{4\pi G D_\mathrm{d} D_\mathrm{ds}} \quad , \tag{2.9b}$$

where $D_\mathrm{ds}$, $D_\mathrm{s}$ and $D_\mathrm{d}$ are the distances from the lens and the observer to the source, and from the observer to the lens, respectively. In terms of the deflection potential

$$\psi(\boldsymbol{\theta}) = \frac{1}{\pi}\int_{\mathrm{I\!R}^2}\mathrm{d}^2\theta'\,\kappa(\boldsymbol{\theta}')\,\ln|\boldsymbol{\theta}-\boldsymbol{\theta}'| \quad , \tag{2.10}$$

we have

$$\begin{aligned}
\boldsymbol{\alpha}(\boldsymbol{\theta}) &= \nabla\psi(\boldsymbol{\theta}) \quad , \\
\kappa(\boldsymbol{\theta}) &= \frac{1}{2}\nabla^2\psi(\boldsymbol{\theta}) \quad , \\
\gamma_1(\boldsymbol{\theta}) &= \frac{1}{2}\left(\psi_{,11}-\psi_{,22}\right) \quad , \\
\gamma_2(\boldsymbol{\theta}) &= \psi_{,12} \quad ,
\end{aligned} \tag{2.11}$$

where indices separated by a comma denote partial derivatives with respect to $\theta_i$. For details concerning these lensing relations, cf. Schneider, Ehlers & Falco (1992, henceforth SEF).

For the image of a source, we can define in analogy to (2.2) the matrix of second moments $Q_{ij}$; due to the invariance of the surface brightness in gravitational light deflection, it can be easily shown that the transformation between the quadrupole matrix of source and image is given as

$$Q^{(\mathrm{s})} = A\,Q\,A \quad , \tag{2.12}$$

where $A$ is the Jacobi matrix of the lens equation, as defined in (2.5). Defining in analogy to (2.4a) the ellipticity $\chi$ of the image, the transformation between $\chi^{(s)}$ and $\chi$ is obtained from (2.12) as

$$\chi^{(s)} = \frac{2(1-\kappa)\gamma + (1-\kappa)^2\chi + \gamma^2\chi^*}{(1-\kappa)^2 + |\gamma|^2 + 2(1-\kappa)\mathcal{R}e\,(\gamma\chi^*)} \quad , \tag{2.13}$$

where $\gamma = \gamma_1 + i\gamma_2$ is the complex shear, and an asterisk denotes complex conjugation. It can be seen immediately that the transformation between $\chi^{(s)}$ and $\chi$ does not depend on $\kappa$ and $\gamma$ separately, but only on the combination

$$g = \frac{\gamma}{(1-\kappa)} \quad , \tag{2.14}$$

with which (2.13) becomes

$$\chi^{(s)} = \frac{2g + \chi + g^2\chi^*}{1 + |g|^2 + 2\mathcal{R}e\,(g\chi^*)} \quad . \tag{2.15a}$$

The inverse transformation is

$$\chi = \frac{\chi^{(s)} - 2g + g^2\chi^{(s)*}}{1 + |g|^2 - 2\mathcal{R}e\,(g\chi^{(s)*})} \quad . \tag{2.15b}$$

We note that the dimensionless surface mass density $\kappa$, and thus the lens parameter $g$, depends on the redshift of the lens and the source.

## 3 Local observables

### 3.1 Basic considerations

All cluster inversion methods are based on the assumption that the sources are oriented randomly on the sky, so that any net orientation of images is due to the tidal effects of the deflecting mass distribution. Given the validity of this assumption, the determining equation simply reads

$$\left\langle \chi^{(s)} \right\rangle = 0 \quad , \tag{3.1}$$

where the average is taken over an ensemble of sources. Throughout this paper we assume that all sources are at the same redshift, so that the lens parameter $g$ is the same for all sources; this assumption is not a very realistic one in general, but may apply approximately for clusters with low redshift. Then, if we assume that $g$ varies little over a region in the lens plane in which $N$ images are seen, we obtain an estimate for $g$ by solving the equation

$$\left\langle \chi_i^{(s)}[g] \right\rangle \equiv \frac{1}{N}\sum_{i=1}^N \chi_i^{(s)}[g] = 0 \quad , \tag{3.2}$$

where we have explicitly written the dependence of the function $\chi^{(s)}$ from (2.15a) on $g$. However, (3.2) does not have a unique solution $g$, as can be seen from the relation

$$\chi^{(s)}[g] = \frac{g}{g^*}\chi^{(s)*}[1/g^*] \quad ; \tag{3.3}$$

hence, if $g$ is a solution of (3.2), so is $1/g^*$. One therefore defines the *(complex) distortion*

$$\delta = \frac{2g}{1+|g|^2} = \frac{2(1-\kappa)\gamma}{(1-\kappa)^2+|\gamma|^2} \quad , \tag{3.4}$$

which is unvariant under the transformation $g \to 1/g^*$, and tries to find an equation for determining $\delta$. Such is obtained by considering

$$\chi_+^{(s)} := \chi^{(s)}[g] + \chi^{(s)}[1/g^*] = \frac{2f + \chi + \chi^*\delta/\delta^*}{1 + \mathcal{R}e\,(\delta\chi^*)} \quad ,$$

$$\chi_-^{(s)} := \chi^{(s)}[g] - \chi^{(s)}[1/g^*] = \left(\frac{1-|g|^2}{1+|g|^2}\right)\frac{\chi - \chi^*\delta/\delta^*}{1+\mathcal{R}e\,(\delta\chi^*)} \quad ;$$

hence, the combination

$$C := \frac{1}{2}\left(\chi_+^{(s)} + \frac{1+|g|^2}{1-|g|^2}\chi_-^{(s)}\right) = \frac{\delta + \chi}{1 + \mathcal{R}e\,(\delta\chi^*)} \tag{3.5}$$

depends solely on $\delta$, and since $\langle \chi^{(s)}[g]\rangle$ and $\langle \chi^{(s)}[1/g^*]\rangle$ both have to vanish, the desired relation for determining $\delta$ becomes

$$N\langle C\rangle = \sum_{i=1}^{N}\frac{\delta + \chi_i}{1 + \mathcal{R}e\,(\delta\chi_i^*)} = 0 \quad . \tag{3.6}$$

Instead of searching the root of equation (3.6), one can solve for $\delta$ by the iteration equation

$$\delta_{n+1} = -\left\langle\frac{\chi}{1+\mathcal{R}e\,(\delta_n\chi^*)}\right\rangle\left(\left\langle\frac{1}{1+\mathcal{R}e\,(\delta_n\chi^*)}\right\rangle\right)^{-1} \tag{3.7}$$

and start the iteration with the linear approximation

$$\delta_0 = -\langle\chi\rangle \quad . \tag{3.8}$$

In fact, the iteration quickly converges to a unique solution. Note that $|\delta| \leq 1$. In terms of $\delta$, the complex shear becomes

$$\gamma = \frac{(1-\kappa)}{\delta^*}\left(1 \pm \sqrt{1-|\delta|^2}\right) \quad , \tag{3.9}$$

where the '+'-sign ('−'-sign) applies for $(1-\kappa)^2 < |\gamma|^2$ $((1-\kappa)^2 > |\gamma|^2)$, or $\det A < 0$ $(\det A > 0)$, respectively.

It is instructive to investigate the behaviour of $\chi^{(s)}$ under the transformation $g \to 1/g^*$. The components of $\chi^{(s)}$ parallel and perpendicular to $\delta$ are

$$\begin{aligned}\chi_\|^{(s)} &\equiv \frac{\mathcal{R}e\,\left(\chi^{(s)}\delta^*\right)}{|\delta|} = \frac{1}{1+\mathcal{R}e\,(\delta\chi)}\left(|\delta| + \frac{\mathcal{R}e\,(\delta\chi^*)}{|\delta|}\right) \quad ,\\ \chi_\perp^{(s)} &\equiv \frac{\mathcal{I}m\,\left(\chi^{(s)}\delta^*\right)}{|\delta|} = \frac{\mathcal{I}m\,(\delta^*\chi)}{1+\mathcal{R}e\,(\delta\chi^*)}\frac{1}{|\delta|}\left(\frac{1-|g|^2}{1+|g|^2}\right) \quad .\end{aligned} \tag{3.10}$$

From these relations, we see that the component parallel to $\delta$ remains unchanged, and the component perpendicular to $\delta$ changes sign, under the transformation $g \to 1/g^*$. In fact, *as long as we consider all sources to have the same redshift, the degeneracy between $g$ and $1/g^*$ cannot be broken by any local measurement.* To see this, we define the probability distribution $p_{\rm s}\left(\chi^{(\rm s)}\right)$ such that $p_{\rm s}\left(\chi^{(\rm s)}\right)\,{\rm d}^2\chi^{(\rm s)}$ is the probability that the source has an ellipticity within ${\rm d}^2\chi^{(\rm s)}$ of $\chi^{(\rm s)}$. Due to the assumed isotropy of the source orientation, this probability distribution can depend only on $\left|\chi^{(\rm s)}\right|$; furthermore, $p_{\rm s}$ is normalized,

$$2\pi \int_0^1 {\rm d}\left|\chi^{(\rm s)}\right|\,\left|\chi^{(\rm s)}\right|\,p_{\rm s}\left(\left|\chi^{(\rm s)}\right|\right) = 1 \quad. \tag{3.11}$$

The probability distribution $p(\chi)$ of the images is then

$$p(\chi) = p_{\rm s}\left(\left|\chi^{(\rm s)}(\chi,g)\right|\right)\left|\frac{\partial^2 \chi^{(\rm s)}}{\partial^2 \chi}\right| \quad;$$

from (3.10) one finds that

$$\left|\chi^{(\rm s)}\right|^2 = 1 - \frac{\left(1-|\delta|^2\right)\left(1-|\chi|^2\right)}{\left[1+{\cal R}{\rm e}\left(\delta\chi^*\right)\right]^2} \quad, \tag{3.12}$$

and the Jacobian of the transformation is readily evaluated; hence,

$$p(\chi) = p_{\rm s}\left(\sqrt{1 - \frac{\left(1-|\delta|^2\right)\left(1-|\chi|^2\right)}{\left[1+{\cal R}{\rm e}\left(\delta\chi^*\right)\right]^2}}\right)\frac{\left(1-|\delta|^2\right)^{3/2}}{\left[1+{\cal R}{\rm e}\left(\delta\chi^*\right)\right]^3} \quad, \tag{3.13}$$

and we see that this probability distribution depends solely on $\delta$, and is thus invariant under the transformation $g \to 1/g^*$. Thus, the best one can hope to measure from the local distribution of galaxy images is the value of $\delta$.

### 3.2 Statistics of single images

From a single galaxy image, or its ellipticity $\chi$, can one obtain some estimate (however uncertain) of $\delta$, or can, for an assumed value of $\delta$, a measure be given for the deviation of $\chi$ from its expectation value? Unfortunately, the answer to the first question is negative: the way one would estimate the expectation value of $\delta$ is to invert (2.15) and write $\delta$ as a function of $\chi$ and $\chi^{(\rm s)}$; then, an expectation value for $\delta$ could be obtained from the probability distribution $p_{\rm s}$. However, for given $\chi$ and $\chi^{(\rm s)}$, there are in general two values of $\delta$ which satisfy the transformation equation (2.15). Thus we are left with the second of the above mentioned possibilities, to determine the statistical properties of $\chi$ for a given value of $\delta$.

The reason why one is interested in these statistical quantities is that of fitting a mass model to observational data. Any mass model predicts $\delta$ as a function of position in the lens plane; this means that for any given galaxy image, one can compare the observed value of $\chi$ with the local expectation value $\langle\chi\rangle$ and define a local error measure $|\chi - \langle\chi\rangle|^2/\sigma^2(\chi)$; the parameter of the lens model can then be chosen as to minimize the sum of the local errors. Here, $\sigma(\chi)$ is the dispersion of $\chi$, which is a function of $|\delta|$. The functions $\langle\chi\rangle$ and $\sigma(\chi)$ depend, besides on $\delta$, also on the intrinsic distribution of source ellipticities. Hence, the usefulness of these measures depends on the degree to which this distribution function can be obtained from observations.

The expectation value $\langle\chi\rangle$ of $\chi$ must have the same phase (or 'direction') as $\delta$; we can therefore consider the special case of $\delta_2 \equiv \mathcal{I}\mathrm{m}(\delta) = 0$, which implies, from (3.4), that $g_2 \equiv \mathcal{I}\mathrm{m}(g) = 0$. The transformation (2.15) then becomes

$$\chi_1 \equiv \mathcal{R}\mathrm{e}(\chi) = \frac{1}{1 - \delta\chi_1^{(\mathrm{s})}} \left(\chi_1^{(\mathrm{s})} - \delta\right) \ ,$$
$$\chi_2 \equiv \mathcal{I}\mathrm{m}(\chi) = \frac{\chi_2^{(\mathrm{s})}}{1 - \delta\chi_1^{(\mathrm{s})}} \frac{1 - g^2}{1 + g^2} \ , \tag{3.14}$$

with $\delta \equiv \delta_1$, $g \equiv g_1$. The expectation values can now be calculated from the probability distribution $p_\mathrm{s}$ of the source ellipticities; after performing the angular integration, we find

$$\langle\chi_1\rangle = 2\pi \int_0^1 \mathrm{d}y \ y \, p_\mathrm{s}(y) \left(\frac{1 - \delta^2}{\delta\sqrt{1 - \delta^2 y^2}} - \frac{1}{\delta}\right) \ ,$$

$$\langle\chi_2\rangle = 0 \ ,$$

$$\langle\chi_1^2\rangle = 2\pi \int_0^1 \mathrm{d}y \ y \, p_\mathrm{s}(y) \left[\frac{\left(1 - \delta^2 y^2\right)^2 - \sqrt{1 - \delta^2 y^2}\left(1 - \delta^4 - 2\delta^2 y^2 + 2\delta^4 y^2\right)}{\delta^2 \left(1 - \delta^2 y^2\right)^2}\right] \ ,$$

$$\langle\chi_1\chi_2\rangle = 0 \ ,$$

$$\langle\chi_2^2\rangle = 2\pi \frac{1 - \delta^2}{\delta^2} \int_0^1 \mathrm{d}y \ y \, p_\mathrm{s}(y) \left[\frac{\left(1 - \delta^2 y^2\right)^2 - \sqrt{1 - \delta^2 y^2}\left(1 - 2\delta^2 y^2\right)}{\left(1 - \delta^2 y^2\right)^2}\right] \ .$$
$$\tag{3.15}$$

For an arbitrary phase of $\delta$, we define the function $\zeta(|\delta|)$ by

$$\langle\chi\rangle = -\zeta(|\delta|)\,\delta \ , \tag{3.16}$$

which in our case ($\delta_2 = 0$) becomes $\zeta(\delta) = -\langle\chi_1\rangle/\delta$. The dispersion of $\chi$,

$$\sigma_\chi = \sqrt{\langle|\chi|^2\rangle - |\langle\chi\rangle|^2} = \sqrt{\langle\chi_1^2\rangle + \langle\chi_2^2\rangle - \langle\chi_1\rangle^2} \ , \tag{3.17}$$

where the final expression applies to our special choice $\delta_2 = 0$, is invariant against rotations of $\delta$. For any given choice of the distribution function $p_\mathrm{s}$, one can now calculate $\zeta(|\delta|)$ and the corresponding dispersion.

Expanding the integrands in (3.15) in powers of $y$, we can obtain an expansion of the mean values in moments of the ellipticity distribution function, defined as

$$M_n = 2\pi \int_0^1 \mathrm{d}y \ y^{(n+1)} p_\mathrm{s}(y) \ ; \tag{3.18}$$

we then obtain:

$$\zeta(\delta) = 1 - \left(1 - \delta^2\right) \left[ \frac{M_2}{2} + \frac{3M_4}{8}\delta^2 + \frac{5M_6}{16}\delta^4 + \mathcal{O}\left(\delta^6\right) \right] \quad ,$$

$$\langle \chi_1^2 \rangle = \delta^2 + \left(1 - \delta^2\right) \left[ \frac{1 - 3\delta^2}{2}M_2 + \frac{3\delta^2\left(3 - 5\delta^2\right)}{8}M_4 + \frac{5\delta^4\left(5 - 7\delta^2\right)}{16}M_6 + \mathcal{O}\left(\delta^6 M_8\right) \right] \quad ,$$

$$\langle \chi_2^2 \rangle = \left(1 - \delta^2\right) \left[ \frac{M_2}{2} + \frac{9M_4}{8}\delta^2 + \frac{25M_6}{16}\delta^4 + \mathcal{O}\left(\delta^6 M_8\right) \right] \quad ,$$

(3.19)

so that

$$\sigma_\chi(\delta) = \sqrt{M_2}\left[ 1 - \frac{6M_2 + M_2^2 - 9M_4}{8M_2}\delta^2 \right.$$
$$\left. - \frac{4M_2^2 - 20M_2^3 + M_2^4 + 108M_2 M_4 + 6M_2^2 M_4 + 81M_4^2 + 200M_2 M_6}{128M_2^2}\delta^4 + \mathcal{O}\left(\delta^6\right) \right] .$$

(3.20)

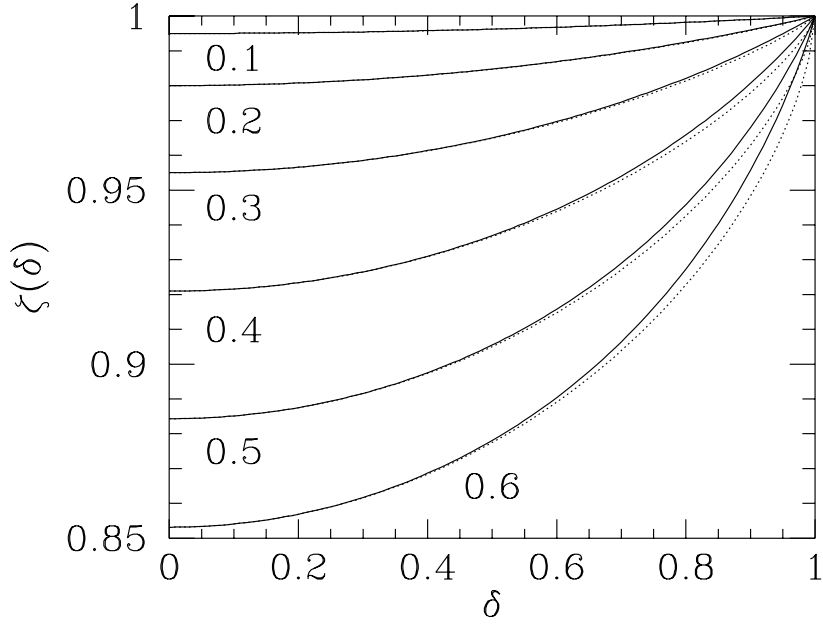

**Fig. 1.** The function $\zeta(\delta)$, defined in (3.16), for an ellipticity distribution of the form (3.21), for several values of the width $R$. The solid curves show the exact value of $\zeta(\delta)$, whereas the dotted curves show the approximation $\bar{\zeta}(\delta)$, see (3.22). It can be seen that $\bar{\zeta}$ provides a good approximation for all values of $\delta$ and $R$. The curves are labelled by the value of $R$

To illustrate the preceding calculation, we consider a specific distribution of source ellipticities, of the Gaussian form

$$p_s\left(\left|\chi^{(s)}\right|\right) = \frac{1}{\pi R^2 \left(1 - e^{-1/R^2}\right)} e^{-\left|\chi^{(s)}\right|^2/R^2} \quad , \tag{3.21}$$

so that $R$ describes the width of the distribution. Its first few moments are

$$M_0 = 1 \quad, \quad M_2 = R^2 - \frac{1}{e^{1/R^2} - 1} \quad,$$

$$M_4 = 2R^4 - \frac{1 + 2R^2}{e^{1/R^2} - 1} \quad, \quad M_6 = 6R^6 - \frac{1 + 3R^2 + 6R^4}{e^{1/R^2} - 1} \quad.$$

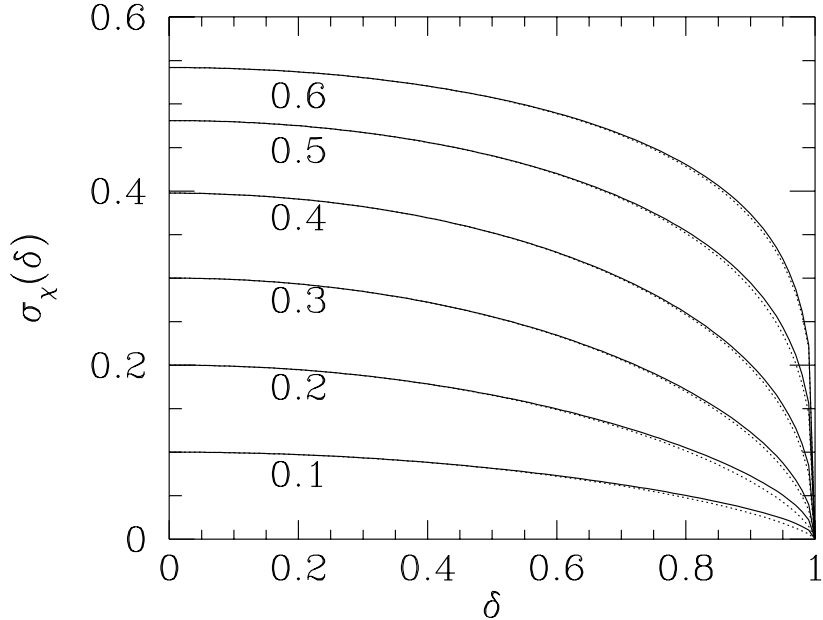

**Fig. 2.** The dispersion $\sigma_\chi$ as a function of $\delta$, for an ellipticity distribution of the sources given by (3.21), for several values of $R$. The solid curves show the exact values of $\sigma_\chi$, the dotted curves are the approximation $\bar\sigma_\chi$, given by Eq.(3.23). The curves are labelled by the value of $R$

In Fig. 1 we have plotted the function $\zeta(\delta)$ (solid curves) for several values of the width $R$. For $\delta \to 1$, $\zeta \to 1$, as can also be seen directly from the the definition (3.16) of $\zeta$ and the expression (3.15) for $\langle\chi\rangle$. For smaller values of $\delta$, $\zeta < 1$, which implies that the linear relation $\delta = -\langle\chi\rangle$ – see (3.8) – *underestimates* the true value of $\delta$. The deviation of $\zeta$ from unity, and thus the error made by using the linear relationship, depends on the width of the ellipticity distribution. Fig. 2 displays the dispersion $\sigma_\chi$ as a function of $\delta$, again for several values of the width $R$ of the distribution (3.21). For $\delta \to 1$, $\sigma_\chi$ vanishes – all sources (formally) attain axis ratio zero if $\delta \to 1$, independent of their intrinsic ellipticity. As expected, the dispersion depends sensitively on the width of the distribution $p_s$. Analogous calculations can be found in ME91.

Since we shall use the functions $\zeta$ and $\sigma_\chi$ extensively below, it will be useful to find an approximation for these function. It turns out that a useful approximation is

$$\bar\zeta(\delta) = 1 - \frac{M_2}{2} \left(1 - \delta^2\right)^{\mu_1} \quad, \qquad (3.22a)$$

with

$$\mu_1 = 1 - \frac{3M_4}{4M_2} \quad, \qquad (3.22b)$$

where the parameters have been fixed by requiring that $\bar\zeta = \zeta$ for $\delta = 0$ and 1, and that the second derivative of both functions at $\delta = 0$ agree. Correspondingly,

$$\bar\sigma_\chi = A\left(1-\delta^2\right)^{\mu_2} \quad , \tag{3.23a}$$

with

$$A = \sqrt{M_2} \quad , \quad \mu_2 = \frac{6M_2 + M_2^2 - 9M_4}{8M_2} \quad , \tag{3.23b}$$

and the parameters have been fixed by the same requirements as those for $\bar\zeta$. In Figs. 1,2, the functions $\bar\zeta$ and $\bar\sigma_\chi$ have been plotted as dotted curves; it is seen that they agree well with $\zeta$ and $\sigma_\chi$, respectively, except if one comes close to $\delta \approx 1$. The special form of the approximations (3.22 & 23) may be due to the specific choice of the intrinsic ellipticity distribution (3.21), but similar functional relationships are expected also for other distributions $p_{\rm s}$.

The measurement of the absolute value of $\chi$ is affected by seeing, noise, tracking problems and other effects. As argued by Kochanek (1990), one should try to use only the orientation of $\chi$ to determine the cluster mass profile, since it is less affected by observational problems that $\chi$ itself. We therefore consider next the statistical properties of the orientation of $\chi$.

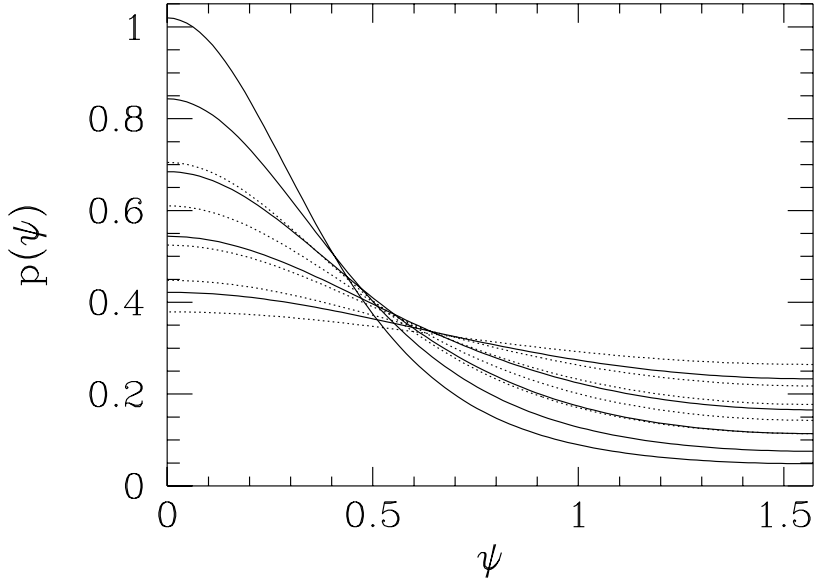

**Fig. 3.** The probability distribution for the angle $\psi$ between the direction of the minor axis of the equivalent image ellipse and the local direction of the shear. The intrinsic ellipticity distribution (3.21) has been assumed, the solid curves are for $R = 0.3$, the dotted curves for $R = 0.5$. The distributions are displayed for $|\delta| = 0.05, 0.1, 0.15, 0.2, 0.25$, with the smaller values corresponding to the flatter distributions

For this, we define $\psi$ to be the angle between the *minor* axis of the equivalent image ellipse, and the local direction of $\delta$: if we write $\chi = \langle\chi\rangle\,{\rm e}^{2{\rm i}\theta}$ and $\delta = |\delta|\,{\rm e}^{2{\rm i}\phi}$, then $\psi = \theta - \phi - \pi/2$. In particular, for an intrinsically round source, $\psi = 0$. The probability distribution for $\psi$ is obtained from (3.13) to be

$$p_{\rm a}(\psi) = 2\int_0^1 {\rm d}\chi\,\chi\,p_{\rm s}\left(\sqrt{1 - \frac{\left(1-|\delta|^2\right)\left(1-|\chi|^2\right)}{\left[1-\delta\chi\cos 2\psi\right]^2}}\right)\frac{\left(1-|\delta|^2\right)^{3/2}}{\left[1-\delta\chi\cos 2\psi\right]^3} \quad , \tag{3.24}$$

which is normalized on $\psi \in [-\pi/2, \pi/2]$. In Fig. 3, we have displayed the function $p_\mathrm{a}(\psi)$, assuming the distribution (3.21) for the source ellipticities, for two values of $R$ and several values of $\delta$. Since $p_\mathrm{a}(\psi)$ is an even function of $\psi$, we have plotted only the interval $\psi \in [0, \pi/2]$. As expected, the larger the value of $R$, and the smaller $\delta$, the flatter is the angular distribution function.

The mean value $\langle \psi \rangle = 0$; this is equivalent to saying that the mean orientation of the images immediately yields the local orientation of the distortion, which in turn agrees with the local orientation of the shear. Information about the strength of the distortion is contained in moments of the angular probability distribution. We define

$$\langle \cos 2\psi \rangle = \int_{-\pi/2}^{\pi/2} \mathrm{d}\psi \; p_\mathrm{a}(\psi) \cos 2\psi \tag{3.25a}$$

and the corresponding dispersion

$$\sigma(\cos 2\psi) = \sqrt{\langle \cos^2 2\psi \rangle - \langle \cos 2\psi \rangle^2} \quad, \tag{3.25b}$$

and display both quantities in Fig. 4. The average $\langle \cos 2\psi \rangle$ increases linearly for small $\delta$, with a slope depending on the intrinsic ellipticity distribution, and approaches unity for $\delta \to 1$, as expected. The corresponding dispersion is $1/\sqrt{2}$ at $\delta = 0$ and vanishes as $\delta \to 1$.

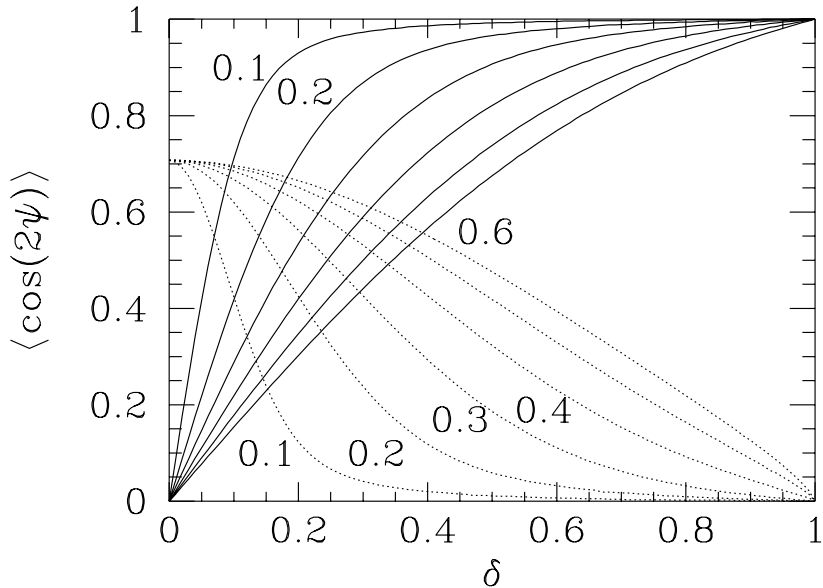

**Fig. 4.** The mean value $\langle \cos 2\psi \rangle$ (solid curves) and the corresponding dispersion $\sigma(\cos 2\psi)$ (dotted curves) as a function of $\delta$, for the choice (3.21) of the intrinsic ellipticity distribution. The curves are labelled by the value of $R$

Unfortunately, there is no simple expression for $\langle \cos 2\psi \rangle$ or $\sigma$ in terms of moments of the intrinsic ellipticity distribution. For later reference, we list here the lowest order term for the intrinsic distribution (3.21),

$$\langle \cos 2\psi \rangle = \frac{\sqrt{\pi}\, \text{erf}(1/R)}{2\left(1 - e^{-1/R^2}\right) R} \delta + \mathcal{O}\left(\delta^3\right) \quad , \tag{3.26}$$

where erf($x$) is the error function. Note that $\langle \cos 2\psi \rangle$ is very insensitive to $\delta$ once this quantity approaches unity; this property will be of some relevance in the next subsection.

### 3.3 Local determination of the distortion

In the previous subsection we have considered the statistical properties of the image ellipticity for a given distortion; here we consider the problem to determine a value of the distortion from a given set of image ellipticities. The situation we have in mind is to have a set of $N$ galaxy images with measured ellipticity $\chi_i$ in an area which is small enough to assume that $\delta$ is appoximately the same for all images. How can $\delta$ be determined?

We consider basically two methods to estimate $\delta$. The first method (hereafter Method 1a) has already been indicated in Sect. 3.1, namely to solve (3.6), the condition which expresses the intrinsic isotropy of the source distribution. The equation is efficiently solved by the iteration (3.7), which after three to five iterations yields the correct solution of (3.6) with an accuracy of better than $10^{-7}$ (i.e., much better than one can ever hope to need). A slight variant of this method is provided by applying the relation (3.16), i.e., to determine the mean ellipticity $\langle \chi \rangle$ from the images and to solve (3.16) for $\delta$, either using the exact form of $\zeta$ or the approximation $\bar{\zeta}$, as given in (3.22) (hereafter Method 1b). The advantage of Method 1a is that the intrinsic ellipticity distribution need not be known, whereas for Method 1b, the function $\zeta(|\delta|)$ depends on the intrinsic ellipticity distribution.

A different method is provided by the statistical properties of the image orientations only, i.e., without using the absolute values of the $\chi_i$. The phase (or direction) of the distribution is given by the relation $\langle \psi \rangle = 0$, or more practically, $\langle \sin 2\psi \rangle = 0$. This yields for the direction $\phi$ of the distortion

$$\tan 2\phi = \frac{\langle \sin 2\theta \rangle}{\langle \cos 2\theta \rangle} \quad . \tag{3.27}$$

Then, the amplitude of the distortion is determined by calculating $\langle \cos 2\psi \rangle$ and (numerically) inverting the function $\langle \cos 2\psi \rangle (\delta)$ plotted in Fig. 4. We henceforth refer to this method as Method 2.

We have run simulations in which $N$ sources were distributed, according to the intrinsic distribution (3.21), and mapped through a distortion $\delta$. From the images, i.e., the $\chi_i$, we have determined the estimates $\delta^{(x)}$, with $x = 1a, 1b, 2$, with each of the above mentioned methods. Repeating the simulation, we have determined the rms error

$$\sigma^{(x)} = \sqrt{\left\langle \left|\delta^{(x)} - \delta\right|^2 \right\rangle} \tag{3.28}$$

for each method, which depends on the distortion $\delta$. These errors are displayed in Fig. 5.

From the statistical results for single images, we can obtain a simple approximation of the rms error corresponding to Method 1. Assuming Poisson statistics, the expected error is $\sigma^{(1)} = \sigma_\chi / \sqrt{N}$, where $\sigma_\chi$ is defined in (3.17) and plotted in Fig. 2. The dotted curves in Fig. 5 correspond to this Poisson error, with $\sigma_\chi$ replaced by $\bar{\sigma}_\chi$, as given in (3.23). We see that this analytical estimate yields a very accurate description of the true error of Method 1. The small, but systematic deviations of the analytical estimate from the true error are due to the fact that $\sigma_\chi$ describes the dispersion of $\langle \chi \rangle$ for a given $\delta$, which is not the same as the dispersion of $\delta$ for a given $\langle \chi \rangle$.

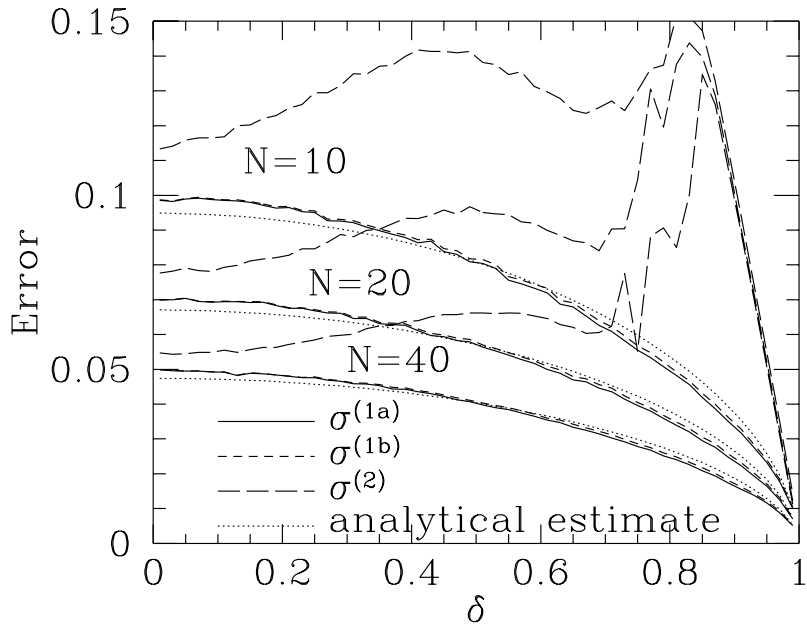

**Fig. 5.** The rms error of the determination of the distortion, as defined in (3.28). These curves were determined from $10^4$ simulations for each value of $\delta$, with $N = 10$, 20, and 40 sources in each case. The intrinsic ellipticity distribution (3.21) was assumed, with $R = 0.3$. The solid curves show the error $\sigma^{(1a)}$ as obtained from Method 1a, short-dashed curves show $\sigma^{(1b)}$, and long-dashed curves show $\sigma^{(2)}$. The dotted curves show the analytic estimate of the error for Method 1, assuming Poisson statistics (see text)

We see from Fig. 5 that in the 'linear' regime, $\delta \ll 1$, the determination of $\delta$ from Method 2 is not very inferiour to that of Method 1, whereas the error is very much larger in the nonlinear regime ($\delta \not\ll 1$). The origin of this large error in the nonlinear regime can be seen from Fig. 4. Since the amplitude of $\delta$ is determined from inverting $\langle \cos 2\psi \rangle$ as a function of $\delta$, and since this function becomes very flat in the nonlinear regime, the amplitude of $\delta$ can only be determined crudely. This, however, does not necessarily imply that the determination of $\delta$ from image orientation alone is precluded. It is likely that one can define statistical quantities from the angular probability distribution function (3.24) which are more sensitive functions of $\delta$ in the nonlinear regime. For example, one could use the function $\langle 1 - 4(\psi - \pi/2)^2/\pi^2 \rangle$ to determine $\delta$ in the nonlinear regime, or make use of the full angular distribution $p_a(\psi)$; we hope to return to these issues in a later paper.

From the numerical simulations, we can also obtain the distribution of estimated values of $\delta$, $p(\delta, \delta_{\text{true}})$, where $\delta_{\text{true}}$ is the true value of $\delta$ with which the galaxy images were imaged. In Fig. 6 we have plotted the contours of equal proability $p(\delta, \delta_{\text{true}})$, for Method 1a and 2. This figure illustrates that the estimated values are isotropically distributed around the true value if $\delta_{\text{true}}$ is in the linear regime, but anisotropies of the error occurs for larger distortions. For Method 1, these anisotropies are oriented such that the error in the 'radial' component of $\delta$ is smaller than that in the tangential direction; this decrease of the error in the radial direction is directly related to the fact

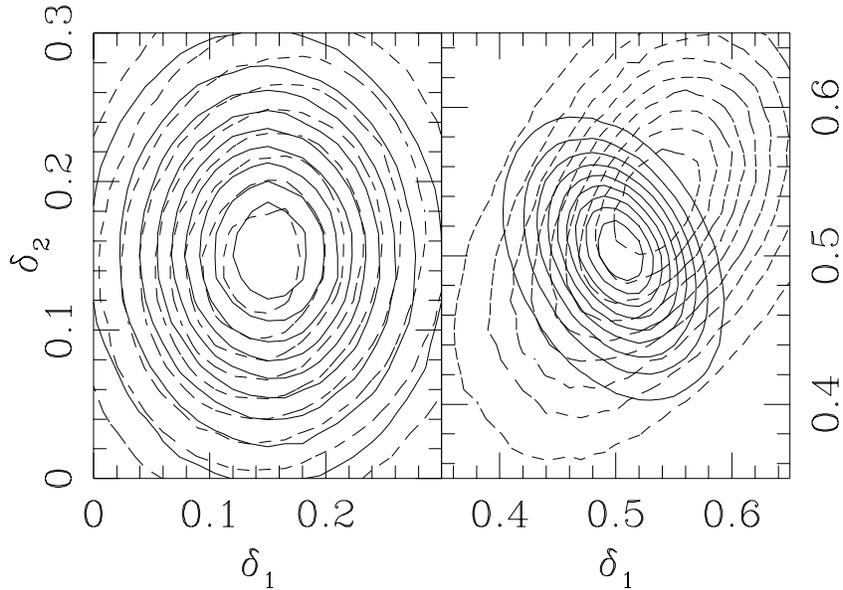

**Fig. 6.** Contours of equal probability $p(\delta, \delta_{\rm true})$ are plotted, for $\delta_{\rm true} = 0.15(1+{\rm i})$ (Panel a) and $\delta_{\rm true} = 0.5(1+{\rm i})$ (Panel b). Solid contours correspond to values of $\delta$ determined by Method 1a, dashed contour to those obtained by Method 2. The levels show contours of 10%, 20%, ..., 90% of the peak probability

that the error of the absolute value of $\delta$ decreases with increasing $|\delta|$, a fact which is illustrated by the decrease of the function $\sigma_\chi$ towards $\delta \to 1$. In contrast, the error in the determination of the direction $\phi$ of $\delta$ from Method 2 is about the same as that from Method 1, but the error of the absolute value of $\delta$ is much larger than that for Method 2, a direct consequence of the flatness of the function $\langle \cos 2\psi \rangle$ for $\delta \to 1$.

### 3.4 An invariance transformation

Can one reconstruct the surface mass distribution $\kappa(\boldsymbol{\theta})$ from the observable image distortion $\delta(\boldsymbol{\theta})$? This question has been investigated by KS in the linear regime, i.e., for the case that one can set $\delta \approx 2g \approx 2\gamma$. They have shown that the finite region of the sky in which the distortion is measured leads to a free parameter in the reconstructed surface mass density, namely an overall additive constant, which is related to the surface mass density at the 'edge of the frame'. Here we show that a similar invariance transformation exists in the fully nonlinear case.

Let $\delta(\boldsymbol{\theta})$ be the observed distortion field, and suppose that the mass distribution $\kappa_0(\boldsymbol{\theta})$ is an inversion of the profile, i.e., $\kappa_0(\boldsymbol{\theta})$ satisfies the relation (3.4). Then, the one-parameter family of mass profiles

$$\kappa(\boldsymbol{\theta}; \lambda) := 1 - \lambda + \lambda \kappa_0(\boldsymbol{\theta}) \qquad (3.29)$$

also yields an inversion of (3.4). This transformation corresponds to an additive constant of the surface mass density, plus a related scaling of $\kappa_0$, and is similar to the 'magnification transformation' pointed out by Gorenstein, Falco & Shapiro (1988; see also Sect. 13.1.1 of SEF). The proof of the validity of this invariance property is simple: first, the transformation (3.29) yields $(1 - \kappa(\boldsymbol{\theta})) = \lambda(1 - \kappa_0(\boldsymbol{\theta}))$. Second, the constant term in (3.29) leads to a linear term in the deflection angle, so that $\boldsymbol{\alpha}(\boldsymbol{\theta}) = (1 - \lambda)\boldsymbol{\theta} + \lambda \boldsymbol{\alpha}_0(\boldsymbol{\theta})$, where $\boldsymbol{\alpha}_0(\boldsymbol{\theta})$ is the deflection angle corresponding to $\kappa_0$. Calculating the Jacobi matrix (2.6) from this deflection law leads to the result that $\gamma = \lambda \gamma_0$. Hence, both the numerator and denominator of (3.4) are multiplied by $\lambda^2$ under the transformation (3.29), so that $\delta$ remains invariant.

In the linear case, $|\lambda - 1| \ll 1$, and the transformation (3.29) indeed corresponds to adding a constant surface mass density, in agreement with KS. The exact nonlinear transformation is only slightly more complicated. It should be noted that the transformation (3.29) leaves the critical curves unchanged; in addition, the curves of $\kappa(\boldsymbol{\theta}) = 1$ are unchanged. In general, the transformation of the magnification is $\mu(\boldsymbol{\theta}, \lambda) = \lambda^{-2} \mu_0(\boldsymbol{\theta})$.

In practice, one can obtain constraints on $\lambda$; for example, since the surface mass density cannot be negative, $\lambda$ must be chosen such that $\kappa \geq 0$ everywhere. Other kinds of possible constraints are to require that $\kappa$ behaves like a power law of $|\boldsymbol{\theta}|$ for large values of $|\boldsymbol{\theta}|$. For the case of a spherical cluster, we will discuss this invariance in somewhat more detail in Sect. 5 below.

## 4 Spherical clusters

As an illustrative application of our previous results, we want to determine in this section the special case of a cluster with axially-symmetric matter distribution, i.e., $\kappa(\boldsymbol{\theta}) = \kappa(|\boldsymbol{\theta}|)$. In Sect. 4.1 we discuss the theoretical behaviour of the distortion in spherical clusters and choose a particular parametrized lens model. The determination of the cluster center is discussed in Sect. 4.2, where we also investigate the minimum mass a cluster must have to produce a signal in the image distortion. In the next section, we then discuss several methods to reconstruct the density profile from observed galaxy images.

### 4.1 The distortion $\delta$

The direction of the shear in a spherical cluster is always radial: at a position $\boldsymbol{\theta} = (\theta \cos \varphi, \theta \sin \varphi)$, the complex shear is $\gamma_1 + \mathrm{i}\gamma_2 = \Gamma \mathrm{e}^{2\mathrm{i}\varphi}$, where $\Gamma(\theta)$ is a real function, not necessarily positive, which depends only on the angular separation $\theta$ of the point considered from the cluster center, chosen here to be at $\boldsymbol{\theta} = 0$. In the case of a spherical cluster, the deflection angle (2.8) reduces to (see SEF, Sect. 8.1)

$$\alpha(\theta) = \frac{m(\theta)}{\theta} =: \frac{2}{\theta} \int_0^\theta \mathrm{d}\theta \, \theta \, \kappa(\theta) \quad, \qquad (4.1)$$

and the direction of $\boldsymbol{\alpha}$ is colinear to that of $\boldsymbol{\theta}$; in the second step of (4.1), we have defined the dimensionless mass $m(\theta)$ inside a circle of radius $\theta$. To determine the function $\Gamma(\theta)$, it is easiest to consider the Jacobian matrix $A(\boldsymbol{\theta})$ on the $\theta_1$-axis: the matrix element $A_{11}$ there becomes

$$A_{11} = 1 - \kappa + \Gamma = 1 - \frac{d\alpha}{d\theta} \quad ,$$

so that

$$\Gamma(\theta) = \frac{m(\theta)}{\theta^2} - \kappa(\theta) =: \bar\kappa(\theta) - \kappa(\theta) \quad , \tag{4.2}$$

where in the second step we have defined $\bar\kappa(\theta)$, the mean dimensionless surface mass density inside a circle of radius $\theta$. For centrally-condensed mass distributions, $\kappa' \equiv d\kappa/d\theta \leq 0$, $\bar\kappa \geq \kappa$, and thus $\Gamma \geq 0$ for all $\theta$. The inversion of (4.2) can be easily obtained by multiplication with $\theta^2$ and differentiation:

$$\kappa(\theta) = -\Gamma(\theta) + 2 \int_\theta^\infty \frac{d\theta'}{\theta'} \Gamma(\theta') \quad , \tag{4.3}$$

where the condition that $\kappa \to 0$ and $\Gamma \to 0$ as $\theta \to \infty$ has been incorporated in the upper integration limit. Equation (4.3) can also be derived from the inversion formula of KS, if the latter is specialized to symmetric matter distributions. This relation provides the basis of cluster inversion methods: if the shear $\Gamma(\theta)$ can be determined observationally, (4.3) can be used to reconstruct the density profile. There are two principal problems the method is faced with: observations will lead to a distribution of galaxy images in a limited solid angle around the cluster, whereas the integral in (4.3) extends to infinity; hence we expect that $\kappa(\theta)$ can be determined from observations of a limited field only up to some transformation (we shall show below how this transformation looks like in the spherical case). Second, as we have discussed in Sect. 3.1, the shear is not an observable; what can be observed is the distortion $\delta$ – see (3.4) – which is related to the shear *and* the local surface mass density. Hence, whereas (4.3) is a linear relation between $\Gamma$ and $\kappa$, the relation between $\delta$ and $\kappa$ is a nonlinear one. In Sect. 5 we shall discuss how a cluster inversion can be performed in the nonlinear case.

We can now discuss the qualitative behaviour of the distortion

$$\delta = \frac{2(1-\kappa)\Gamma}{(1-\kappa)^2 + \Gamma^2} \tag{4.4}$$

as a function of $\theta$, for 'typical' mass profiles. As such we consider profiles $\kappa(\theta)$ which are finite ($\kappa \to 0$ for $\theta \to \infty$), centrally condensed ($d\kappa/d\theta \leq 0$), and which have at most two critical curves, i.e., circles at which the determinant $\det A$ of the Jacobi matrix (2.6) vanishes (note that, in contrast to common belief, even a centrally condensed symmetric lens can have more than two critical curves, but in order for this to occur, the mass profile must be fairly special). From the discussion in SEF, Sect. 8.1.1, we see that critical curves occur if, and only if, the central surface mass density $\kappa(0) \geq 1$. The radius of the tangential (radial) critical curve will be denoted by $\theta_t$ ($\theta_r$), with $\theta_r < \theta_t$, and is given by

$$\bar\kappa(\theta_t) = 1 \quad ; \quad 1 + \bar\kappa(\theta_r) - 2\kappa(\theta_r) = 0 \quad . \tag{4.5}$$

Alternatively, at the critical curves, $1 - \kappa \pm \Gamma = 0$. Furthermore, $\kappa(\theta_t) < 1$, $\kappa(\theta_r) > 1$. Together with the property that $\Gamma \geq 0$ for centrally condensed lenses, this implies that $\delta = +1$ ($\delta = -1$) at the position of the tangential (radial) critical curve, and $|\delta| \neq 1$ elsewhere. Hence, if $\kappa(0) > 1$, $\delta$ will decrease from 0 to $-1$ as $\theta$ increases from 0 to $\theta_r$, then increase from $-1$ to $+1$ for $\theta_r \leq \theta \leq \theta_t$, becoming zero at the point where $\kappa = 1$, and decrease towards 0 as $\theta \to \infty$. If $\kappa(0) < 1$, $\delta$ remains positive but smaller than $+1$ for all $\theta$. Correspondingly, the function

$$g = \frac{\Gamma}{(1-\kappa)} \qquad (4.6)$$

takes values in the range $0 \leq g(\theta) < 1$ if $\kappa(0) < 1$, whereas for $\kappa(0) > 1$, $g$ decreases from 0 to $-1$ as $\theta$ increases from 0 to $\theta_r$, decreases towards $-\infty$ when $\kappa(\theta) = 1$, and then decreases towards zero as $\theta \to \infty$, with $g = 1$ at $\theta_t$. For a mass profile of the form

$$\kappa = \frac{x_c^2 + x^2/2}{(x_c^2 + x^2)^{3/2}} \quad , \qquad (4.7a)$$

so that $\kappa(0) = x_c^{-1}$, with

$$x = \frac{\theta}{\theta_0} \quad , \quad x_c = \frac{\theta_c}{\theta_0} \quad , \qquad (4.7b)$$

where $\theta_c$ is the angular radius of the core of the mass distribution, and $\theta_0$ is a characteristc angular scale, we have plotted the functions $\delta(x)$ and $g(x)$ in Fig. 7.

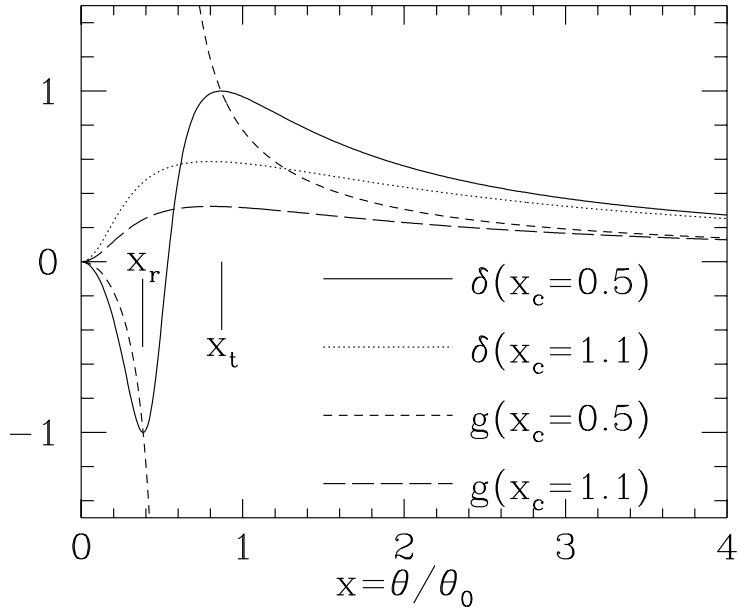

**Fig. 7.** For the mass profile (4.7), the distortion $\delta$ and $g$ are plotted for two different values of the core radius $x_c$. For $x_c = 0.5$, this lens models has two critical curves, the location of which are indicated by $x_r$ and $x_t$ (also in units of $\theta_0$)

Note that positive (negative) values of $\delta$ imply that the image of a round source is aligned tangentially (radially) with respect to the cluster center. At a point $\boldsymbol{\theta} = |\boldsymbol{\theta}|(\cos\phi, \sin\phi)$, we define the *tangential alignment* $\chi_t$ of an image with complex ellipticity $\chi = |\chi| e^{2i\vartheta}$ as

$$\chi_t := -\mathcal{R}e\left(\chi e^{-2i\phi}\right) = -|\chi|\cos[2(\vartheta - \phi)] = |\chi|\cos(2\psi) \quad , \qquad (4.8)$$

where, as before, $\psi = \vartheta - \phi - \pi/2$ is the angle between the radial direction to the cluster center and the minor axis of the image.

## 4.2 Determination of the lens center

Given a set of observed galaxy images mapped through a foreground cluster, one can determine the center of this cluster from the observed image ellipticities. In this subsection, we investigate several methods for this center determination, and also investigate the question of how many galaxy images one needs to detect a signal of lensing to occur.

Consider first the case that the cluster is not critical, i.e., that $\kappa(0) < 1$, so that no critical curves occur. In that case, $\delta \geq 0$ for all $\theta$, so that one expects a net tangential alignment at all radii. For each 'test center' $\mathbf{X}$, one can define the function

$$C_1(\mathbf{X}) = \sum_{i=1}^{N} \cos 2\psi_i(\mathbf{X}) \quad , \tag{4.9a}$$

where the sum extends over all images, and $\psi_i(\mathbf{X})$ is the angle between the direction of a galaxy image to the test center and the minor axis of the image. One expects this function to have a maximum at the true center, chosen here to be at $\mathbf{X} = 0$. In the function $C_1$, each galaxy image attains the same weight, independent of the absolute value of its ellipticity. One might expect that at radii where the distortion is strong, and thus the image ellipticities (or the tangential alignments $\chi_t$) are large, the local direction to the center is better constrained, and thus define

$$C_2(\mathbf{X}) = \sum_{i=1}^{N} \chi_{ti} = \sum_{i=1}^{N} |\chi_i| \cos 2\psi_i(\mathbf{X}) \quad . \tag{4.9b}$$

If we knew the mass profile of the cluster, we could also calculate the function

$$C_3(\mathbf{X}) = \sum_{i=1}^{N} \delta(\mathbf{x}_i) \cos 2\psi_i(\mathbf{X}) \quad , \tag{4.9c}$$

which should be less sensitive to statistical fluctuations than (4.9b): an intrinsically strongly elongated source would be strongly weighted by $C_2$, though the direction of its image will hardly be affected by a weak distortion, whereas the function $C_3$ would give little weight to such an image. Of course, the mass profile is not known a priori, but as will be shown in the nect section, it can be reconstructed, and from this reconstructed profile the function $\delta(x)$ can be calculated and used in the calculation of $C_3$.

If the cluster is critical, i.e., $\kappa(0) > 1$, those galaxy images for which $\kappa > 1$ will cause a negative contribution on average to the sums in (4.9a&b), since they are radially elongated. We therefore define, in analogy with (4.9a-c), the following three functions

$$C_4(\mathbf{X}) = \sum_{i=1}^{N} \cos 4\psi_i(\mathbf{X}) \quad , \tag{4.10a}$$

$$C_5(\mathbf{X}) = \sum_{i=1}^{N} |\chi_i|^2 \cos 4\psi_i(\mathbf{X}) \quad , \tag{4.10b}$$

$$C_6(\mathbf{X}) = \sum_{i=1}^{N} \delta^2(\mathbf{x}_i) \cos 4\psi_i(\mathbf{X}) \quad . \tag{4.10c}$$

It is easily seen that both, tangentially and radially aligned images will contribute positively to the sums in (4.10a-c), so that one might expect that these are better measures to determine the cluster center than the functions (4.9) in the case of critical clusters. One way to visualize the difference between the functions $C_1, C_2$ and $C_4, C_5$ is by noting that $\cos 2\psi_i$ is a function which is positive on the wedge centered on the minor axis of the galaxy image, whereas it is negative on the wedge along the major axis; in contrast, $\cos 4\psi_i$ is positive on all four wedges centered on the axes of the images, and negative in between (see also Kochanek 1990 for a similar geometric argument, applied to the arclets in A370). With this picture in mind, one would expect that the functions (4.10) yield a better 'resolution' of the cluster center than those in (4.9). We shall see below that this expectation is only partly justified. We have performed numerical simulations for the application of the foregoing methods; these simulations will be briefly described next.

We simulated cluster mass distributions of the form (4.7); note that this profile behaves like an isothermal sphere for $x \gg x_c$, $\kappa \to 1/(2x)$, and so the characteristic angular scale $\theta_0$ can be parametrized by the velocity dispersion $\sigma_v$ of an isothermal sphere,

$$\theta_0 = 4\pi \left(\frac{\sigma_v}{c}\right)^2 \left(\frac{D_{\mathrm{ds}}}{D_{\mathrm{s}}}\right) \quad . \tag{4.11}$$

For all simulations presented in this paper, we have set the distance ratio $D_{\mathrm{ds}}/D_{\mathrm{s}} = 0.6$. Galaxies were randomly placed on the lens plane with a density $N_0$ per square arcminute. In principle, one would have to distribute the galaxies randomly in the source plane; however, the source counts of faint galaxies have a alope of close to $-1$, so that the magnification effects are just compensated by the corresponding enlargement of the solid angle. Each galaxy was then assigned an intrinsic ellipticity $\chi^{(s)}$, according to the distribution (3.21). Using the local values of $\kappa$ and $\gamma$, these galaxies were imaged, i.e., their ellipticity $\chi$ was calculated.

Using such simulations, we have determined estimates for the cluster center by maximizing the function $C_i(\mathbf{X})$, $i = 1$ through 6. In Fig. 8 we have plotted the probability of the deviation of the estimated cluster center from the true center, measured in arcminutes, for critical and non-critical clusters, for a circular field of galaxy images with radius $2'.5$ and a galaxy density of $N_0 = 16$; i.e., about 316 galaxies were used for the determination of the cluster center. For the function $C_3$ and $C_6$, the known distribution of $\delta$ was used. It can be seen that the distribution of the deviations of the estimated cluster center from the true one is quite different for the six different functions $C_i$. In fact, for the non-critical clusters, shown in the upper panels of Fig. 8, the functions $C_4$, $C_5$ and $C_6$ are consistently more inaccurate than the functions $C_1$ through $C_3$, containing $\cos 2\psi_i$, whereas for critical clusters, the functions $C_4$ through $C_6$ yield more accurate results than those of (4.9). In all cases, the tendency that $C_3$ is more accurate than $C_2$, which in turn is more accurate than $C_1$, is present, and the same is true for $C_6$, $C_5$ and $C_4$. However, the gain in accuracy by changing from $C_2$ to $C_3$ (and from $C_5$ to $C_6$) is not dramatic, and therefore not really worth the effort, given the additional uncertainty in practice about the functional behaviour of $\delta(\mathbf{x})$. ME91 estimated that the precision with which the cluster center can be determined is of the order of the mean separation of galaxy images; this estimate was based on the simplifying assumption that one uses only those parts of the cluster outside the critical curves. In our case, the mean separation of galaxy images is $0'.25$. The median value of the deviation of the estimated cluster center

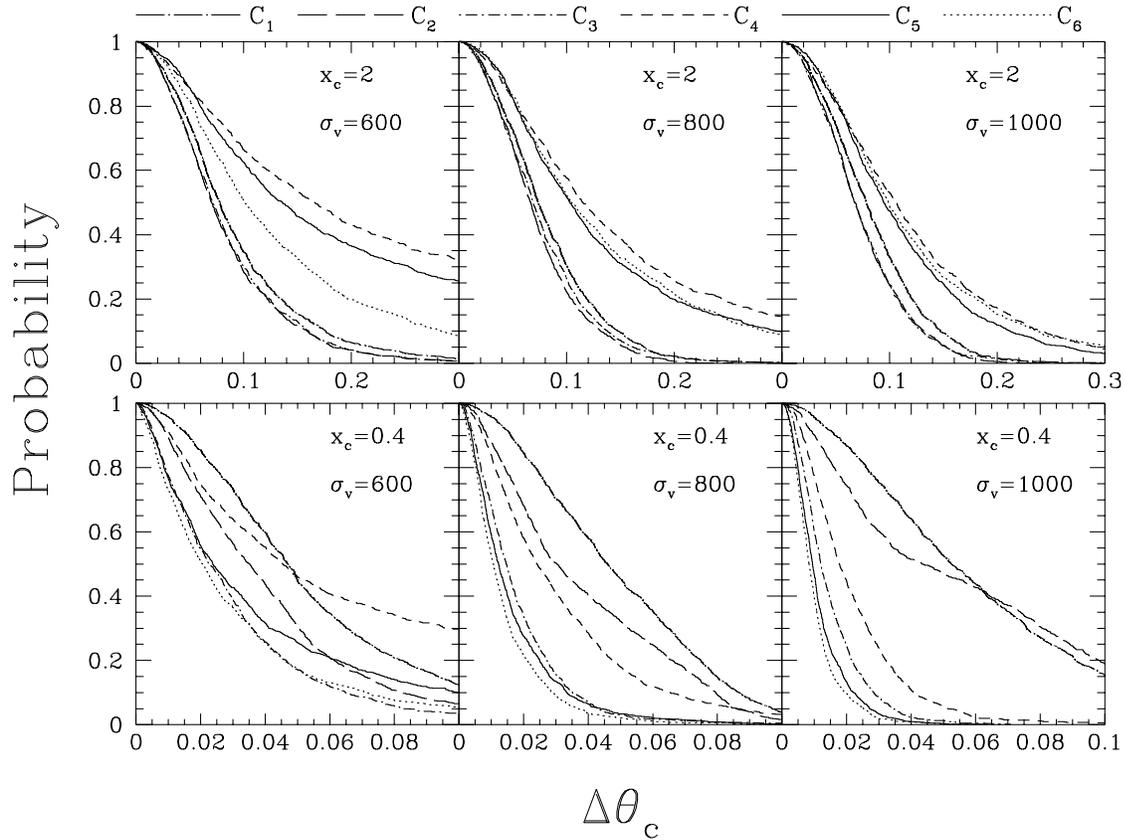

**Fig. 8.** Determination of the cluster center from image distortions. Plotted is the probability $P(\Delta\theta)$ that the error of the estimated cluster center is larger than $\Delta\theta$, measured in arcminutes. For these simulations, a field of radius $2'.5$ was chosen, with mean density $N_0 = 16$ per square arcminute, i.e., the expected number of galaxies per field is 314. The intrinsic ellipticities of the sources were drawn from the distribution (3.21), with $R = 0.3$. The different panels show this probability for different cluster parameters, described by the velocity dispersion $\sigma_v$ and the core radius $x_c$. The upper panels correspond to clusters which are too weak to produce critical curves, whereas the lower panels are for critical clusters with $\kappa(0) > 1$; note the different scales in the upper and lower panels. The center was estimated for 1000 realizations of the galaxy distribution, by maximizing the functions $C_i(\mathbf{X})$ of (4.9 & 10), and the different curve correspond to different functions $C_i(\mathbf{X})$, as indicated at the top of the figure

from the true one is about $0'.08$ as determined by the functions in (4.9) for the non-critical clusters, quite independent of $\sigma_v$, in rough agreement with the estimate of ME91. However, this median deviation becomes considerably smaller for the critical clusters, of order $0'.02$ for the critical clusters, if determined by the functions of (4.10). This is only one tenth of the mean separation of galaxy images. We therefore conclude that the cluster center can be very accurately determined (for assumed spherical clusters) if the cluster is critical and thus capable of producing a few fairly strongly distorted images.

Without determining the mass profile of a cluster, it is not known whether it is singular or not; hence, one would like to obtain a profile-independent criterion on which function $C_i$ yields the most accurate determination of the cluster center. In fact, as we shall see there are two different aspects to this question: one is the accuracy of the center position, and the other is the reliability of the result, or its stability. What we mean by this remark can be best explained by considering the contour levels of the functions $C_1(\boldsymbol{\theta})$ and $C_4(\boldsymbol{\theta})$, as plotted in Fig. 9a, for four different cluster profiles. The dashed contour

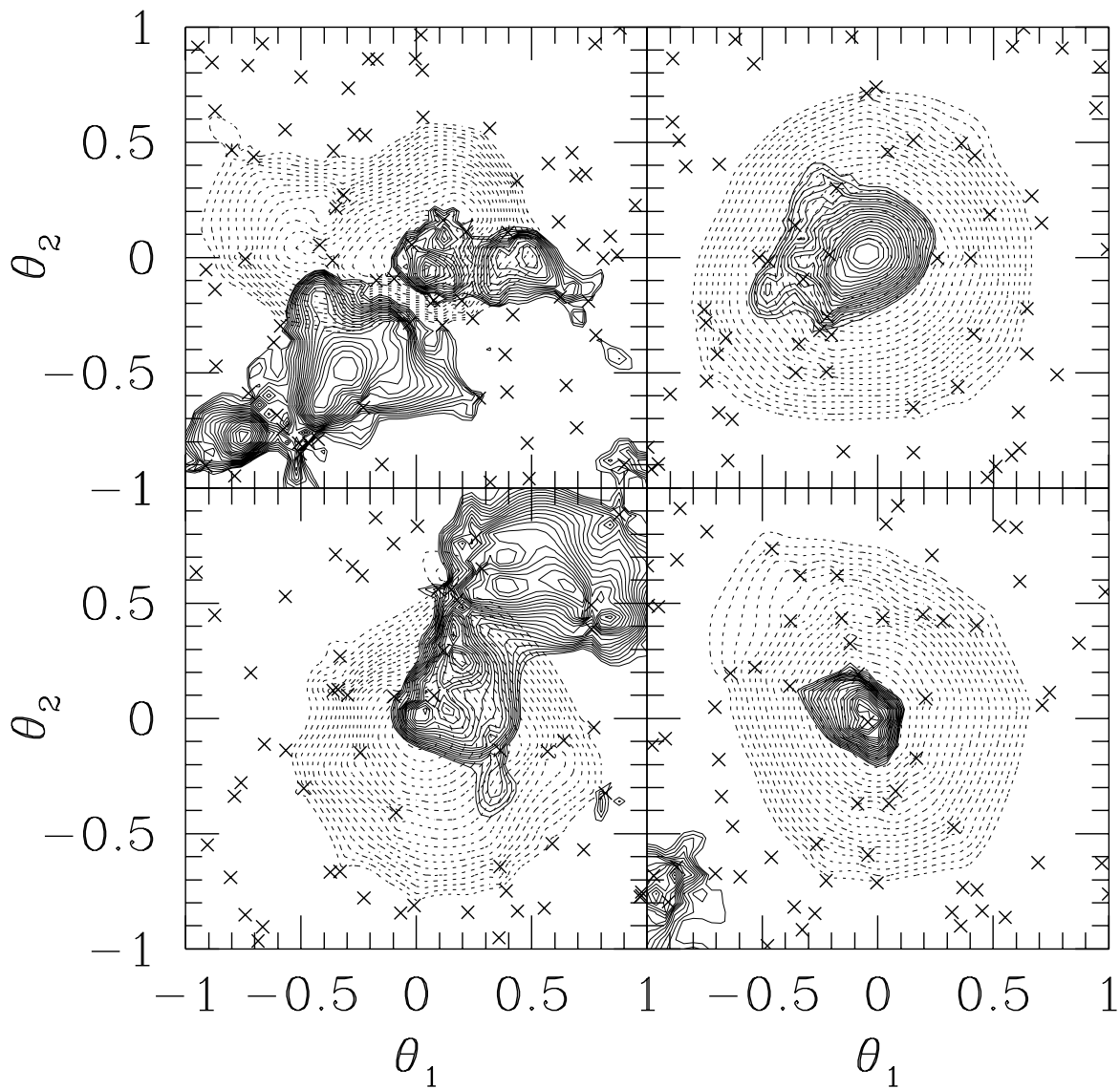

**Fig. 9a.** Contour lines of the function $C_1(\boldsymbol{\theta})$ (dashed contour) and of $C_4(\boldsymbol{\theta})$ (solid contours) for four different clusters. The lensing parameters are the same as those of Fig. 8; the upper left panel has $\sigma_v = 600\,\mathrm{km/s}$, $x_c = 1.5$, the upper right panel has $\sigma_v = 1000\,\mathrm{km/s}$, $R_c = 1.5$; the lower left panel has $\sigma_v = 600\,\mathrm{km/s}$, $x_c = 0.4$, the lower right panel has $\sigma_v = 800\,\mathrm{km/s}$, $x_c = 0.4$. In each case, the inner $2 \times 2$ arcminutes of the cluster are shown, with the true cluster center at $\boldsymbol{\theta} = 0$. The crosses in each case indicate the galaxy images in the central part of the cluster. The levels of the contours are at $60\%, 62\%, \ldots, 98\%$ of the maximum value of the respective function. Note how individual galaxies, or small groups of galaxies, have an impact on the local shape of the contours, in particular for those corresponding to $C_4$

correspond to $C_1$, the solid contours to $C_4$. In all cases displayed, the contours of $C_1$ are fairly round, well concentrated around the true center of the cluster, without having secondary maxima (except perhaps very shallow ones, close to individual galaxy images).

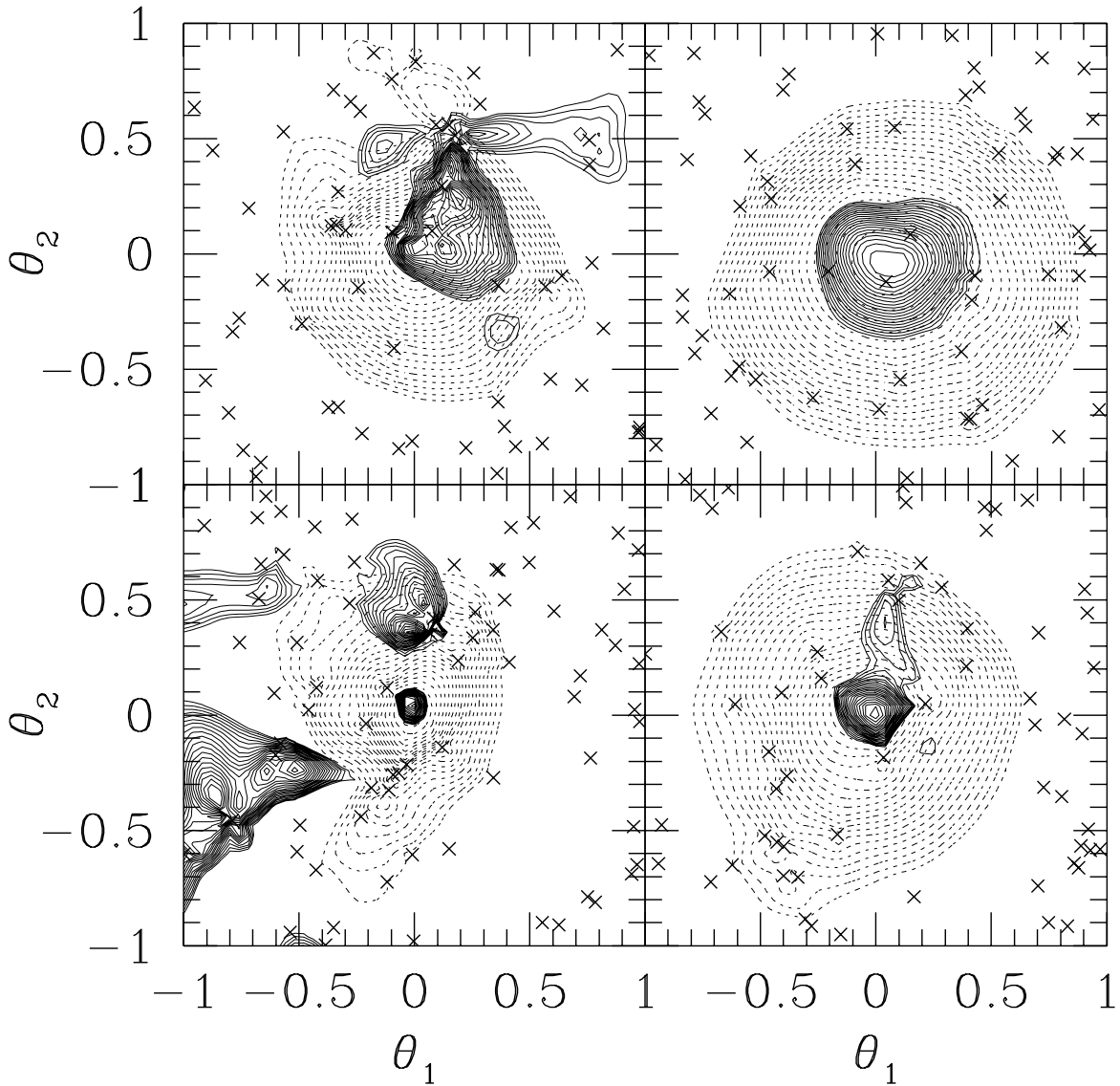

**Fig. 9b.** Same as in Fig. 9a, except that the contour lines of $C_2(\boldsymbol{\theta})$ (dashed contours) and $C_5(\boldsymbol{\theta})$ (solid contours) are plotted. Also, the cluster parameters in the four panels are the same as for Fig. 9a

In contrast, the contours of $C_4$ are much more complicated, in particular for the clusters with small $\sigma_v$. However, for the cluster in the lower right panel of Fig. 9a, the contours of $C_4$ are very well concentrated, and the maximum of $C_4$ lies close to the maximum of $C_1$. By visual inspection of many such contour diagrams, we concluded that the estimate of the center from $C_4$ is nearly always more accurate if the contours are 'simple', without secondary maxima, and located at those positions where the function $C_1$ is not much smaller than its maximum value. On the other hand, in particular for weak clusters, the maximum of $C_4$ can be fairly far away from the true lens center. From this we conclude than the determination of the cluster center is more reliable from $C_1$, but more accurate from $C_4$ if the topology of the contours of $C_4$ are simple and concentrated around the maximum of $C_1$. Unfortunately, we cannot provide a more quantitative criterion on which of the function $C_i$ should be used in a given situation. Since we have seen above that $C_2$ and $C_5$ yield a more accurate center position than $C_1$ and $C_4$, respectively, we have also plotted the contour levels of these two functions in Fig. 9b. Similar remarks as those to Fig. 9a apply here. The function $C_2$ yields a more reliable result than $C_5$, in the sense that the topology of the contours of $C_2$ are 'simple'. The maximum of $C_5$ yields a more accurate determination of the cluster center in those cases where the topology of its contours are simple and centered near the maximum of $C_2$.

Finally, we consider the requirement on the lens strength to produce a significant image distortion for being detected. Suppose that we suspect a mass concentration at a point $\boldsymbol{\theta} = 0$ (e.g., due to the presence of emission there); from observing galaxy images within a circle of radius $\theta_{\rm out}$ around this point, we can detect the mass concentration by comparing the value of the function $C_2(0)$ with the dispersion of this functional value caused by an undistorted distribution of galaxy images. Without a lens, we have

$$\langle C_2 \rangle_0 = 0 \; ; \; \sigma(C_2) = \sqrt{\frac{N}{2}} \sqrt{M_2} \quad , \tag{4.12}$$

where $M_2$ is the mean square ellipticity of the galaxies, as defined in (3.18), and $N = \pi n \theta_{\rm out}^2$ is the number of galaxy images. The mean value of $C_2(0)$ in the presence of a lens is

$$\langle C_2(0) \rangle = \sum_{i=1}^{N} \langle \chi_{{\rm t}i} \rangle \approx 2\pi n \int_0^{\theta_{\rm out}} {\rm d}\theta \; \theta \; \langle \chi_{{\rm t}i} \rangle = 2\pi n \int_0^{\theta_{\rm out}} {\rm d}\theta \; \theta \; \zeta(\delta(\theta)) \, \delta(\theta) \quad , \tag{4.13}$$

where in the final step we made use of (3.16), and the function $\zeta(\delta)$ is plotted in Fig. 1. Hence, in order to detect a significant distortion, we require the "signal-to-noise" ratio

$$\frac{\langle C_2(0) \rangle}{\sigma(C_2)} = \frac{2^{3/2} \sqrt{N}}{\sqrt{M_2} \, \theta_{\rm out}^2} \int_0^{\theta_{\rm out}} {\rm d}\theta \; \theta \; \zeta(\delta(\theta)) \, \delta(\theta) \tag{4.14}$$

to be larger than unity. As an illustrative example, we use a singular truncated isothermal sphere, i.e., we use (4.7a) with $\theta_{\rm c} = 0$ for $\theta \le \theta_{\rm trunc}$, and $\kappa = 0$ for $\theta > \theta_{\rm trunc}$. We then have

$$\begin{aligned} \delta(x) &= \frac{2x-1}{1-2x+2x^2} \quad \text{for } x \le x_{\rm trunc} \\ \delta(x) &= \frac{2x_{\rm trunc} x^2}{x^4 + x_{\rm trunc}^2} \quad \text{for } x > x_{\rm trunc} \quad , \end{aligned} \tag{4.15}$$

with $x_{\rm trunc} = \theta_{\rm trunc}/\theta_0$. Setting $\zeta(\delta) \approx 1$ – see Fig. 1 – and replacing the lower integration limit by $\theta_0$ (i.e., avoiding galaxy images within the critical curve of the mass distribution), we find for $x_{\rm trunc} \gg 1$

$$\frac{\langle C_2(0) \rangle}{\sigma(C_2)} = \frac{2\sqrt{2\pi n}\,\theta_0}{\sqrt{M_2}} f \quad ,$$

where

$$f = \frac{\theta_{\rm trunc}}{\theta_{\rm out}} \left[ 1 + 2\ln\left(\frac{\theta_{\rm out}}{\theta_{\rm trunc}}\right) \right]$$

is a factor of order unity if $\theta_{\rm out}$ is not much larger than $\theta_{\rm trunc}$, but decreases for larger $\theta_{\rm out}$. In other words, increasing the radius within which galaxies are included in the determination of $C_2$ much beyond the truncation radius weakens the signal-to-noise ratio. In useful units, we obtain

$$\frac{\langle C_2(0)\rangle}{\sigma\left(C_2\right)} \approx 4.6 f\left(\frac{D_{\rm ds}}{D_{\rm s}}\right)\left(\frac{\sigma_v}{300{\rm km/s}}\right)^2 \sqrt{\frac{N_0}{40}}\left(\frac{\sqrt{M_2}}{0.3}\right)^{-1} \quad . \tag{4.16}$$

Hence, even isothermal distributions with a velocity dispersion similar to that of massive cD galaxies can be detected from distortions, provided they extend to sufficiently large radii (i.e., several times their critical radius) and the galaxy density is sufficiently large. A similar result has been obtain by ME91.

## 5 Reconstruction of circular cluster profiles

Given a radially-symmetric mass distribution, we want to investigate methods to reconstruct the density profile from distorted galaxy images. In this section, three different methods will be discussed: in Sect. 5.1, the direct inversion will be treated, whereas Sects. 5.2 & 3 deal with fitting methods. Briefly, the latter are the preferred methods if the density of galaxy images is relatively small, whereas the direct inversion yields very accurate profiles for large galaxy density. Throughout this section, we will work in terms of normalized coordinates $x = \theta/\theta_0$, where $\theta_0$ is a reference angle; if specific examples are treated, we shall use the profile (4.7a) and the relation (4.11) between $\theta_0$ and the asymptotic velocity dispersion $\sigma_v$ of the cluster.

### 5.1 Direct inversion

The basis for the inversion of clusters is provided by equations of the form (4.2) or (4.3). For example, combining (4.3) and (4.6), we obtain

$$\kappa(x) = -g(x)\bigl[1 - \kappa(x)\bigr] + 2\int_x^\infty \frac{{\rm d}x'}{x'}\bigl[1 - \kappa(x')\bigr] g(x') \quad .$$

By rearranging terms and differentiation with respect to $x$, this equation can be written in the form

$$\frac{{\rm d}}{{\rm d}x}\bigl[1 - \kappa(x)\bigr]\bigl[1 - g(x)\bigr] = \bigl[1 - \kappa(x)\bigr]\left(\frac{2g(x)}{x} + \frac{{\rm d}g(x)}{{\rm d}x}\right) \quad .$$

This differential equation can be readily solved to yield the inversion

$$|1 - \kappa(x)| = \frac{a}{|1 - g(x)|}\exp\left(-\int_x^\infty {\rm d}x'\,\frac{2g(x')}{x'\bigl[1 - g(x')\bigr]}\right) \quad , \tag{5.1}$$

where $a$ is a constant of integration. The latter one is related to the transformation (3.29), which leads $g(x)$ invariant, but which changes $1 - \kappa(x)$ to $\lambda[1 - \kappa(x)]$. Hence, without additional assumptions (e.g., that the surface mass density tends to 0 as $x \to \infty$), this integration constant cannot be determined from observations of image deformations.

Whereas (5.1) provides an exact reconstruction of the cluster profile in terms of $g(x)$, it is useless for practical calculations. The absolute values imply that one has the freedom to choose between signs. Furthermore, the denominator in (5.1) vanishes at the tangential critical curve, although $1 - \kappa$ remains finite there. This implies that the exponential factor has to compensate for the divergence, which indeed is the case, but such a situation will be highly unstable numerically.

A better way for the inversion is provided by combining (4.2) and (4.6), which yields

$$\kappa(x) = \frac{1}{1 - g(x)} \left( \frac{2}{x^2} \int_0^x dx' \, x' \, \kappa(x') - g(x) \right) \quad . \tag{5.2}$$

As we shall see below, the denominator $(1 - g)$ provides some difficulty for applying (5.2) near critical curves, but this problem can be circumvented. If we assume for a moment that the cluster has no critical curve, then the solution of (5.2) is straightforward. Suppose the function $g(x)$ were given for $x \in [0, x_{\rm out}]$; then, in order to solve (5.2), one has to choose a value $\kappa(x_{\rm out})$, and to calculate

$$m(x_{\rm out}) \equiv 2 \int_0^{x_{\rm out}} dx' \, x' \, \kappa(x') = \bigl([1 - g(x_{\rm out})]\kappa(x_{\rm out}) + g(x_{\rm out})\bigr) x_{\rm out}^2 \quad .$$

Then, by choosing a grid in $x$, with $x_i = i\,\Delta x$, $1 \le i \le N$, $N\,\Delta x = x_{\rm out}$, and approximating the integral in (5.2) by Simpson's rule, one finds the iteration equations

$$\begin{aligned}
\kappa_i &= \frac{m_{i+1} - \Delta x \, x_{i+1} \, \kappa_{i+1} - x_i^2 g_i}{(1 - g_i)x_i^2 + x_i \, \Delta x} \quad , \\
m_i &= m_{i+1} - (x_i \kappa_i + x_{i+1}\kappa_{i+1}) \, \Delta x \quad ,
\end{aligned} \tag{5.3}$$

with $g_i \equiv g(x_i)$ etc. This set of iteration equations can be solved readily for a given function $g(x)$, and it will produce stable results as long as $g$ does not come close to unity. Note that the value of $N$, or, equivalently, $\Delta x$ is at our disposal, so we can make the iterative solution of (5.2) as accurately as desired.

The largest problem is of course that the function $g(x)$ can at best be determined approximately from observations. The observations yield the tangential alignment $\chi_{\rm t}$ of the individual galaxy images; from that, local estimates of the distortion $\delta(x)$ can be obtained. Here, we proceed as follows: for each grid point $x_i$, we average the tangential alignment $\chi_{\rm t}$ of the galaxy images, weighted with a Gaussian; the width of the Gaussian can be adapted to the quality of data, and can also be adjusted to depend on $\langle \chi_{\rm t} \rangle$. This average then yields a first estimate of $\delta_i \equiv \delta(x_i)$. According to (3.16), we then choose $\delta_i = \langle \chi_{\rm t} \rangle / \zeta(\langle \chi_{\rm t} \rangle)$, with the approximation (3.22) for the function $\zeta$. The value of $g_i$ then follows from (3.4) as

$$g_i = \frac{1 \pm \sqrt{1 - \delta_i^2}}{\delta_i} \quad , \tag{5.4}$$

where the minus sign applies for regions in the lens plane where the Jacobian matrix has positive determinant, and vice versa (cf. also the discussion below Eq.(3.9)). The problem with this prescription of $g$ is of course that the value of the determinant is not known a priori; we will discuss the problems related to this further below.

If we assume that the cluster is not critical, the problem of choosing the correct sign in (5.4) does not occur, i.e., we have to take the minus-sign. Fig. 10 shows an example

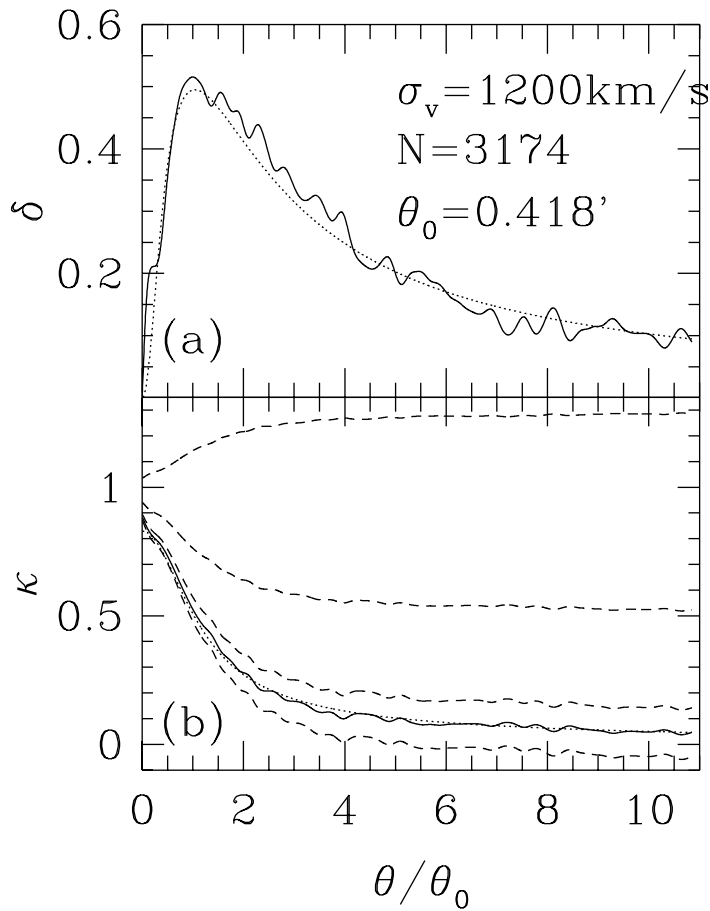

**Fig. 10.** For a cluster with the mass profile (4.7a), corresponding to a velocity dispersion of $\sigma_v = 1200\,\mathrm{km/s}$ and a core radius of $0\rlap{.}'5$, galaxies of density 40/squarearcmin and with internal ellipticity distribution (3.21), with $R = 0.3$ were mapped within the inner 5 arcminutes; in the realization shown here, this amounts to $N = 3174$ galaxies. The cluster is non-critical. (a) The dotted curve shows the theoretical distortion $\delta(x)$, with $x = \theta/\theta_0$, whereas the solid curve is the distortion $\delta$ obtained by smoothing the tangential alignment profile, as described in the text. (b) The dotted curve is the radial mass profile $\kappa(x)$ of the cluster. The solid curve shows the reconstructed mass profile, as obtained from (5.3) and the function $g(x)$, related to the distrotion profile $\delta$ by (5.4). The value of $\kappa(x_{\mathrm{out}})$ was chosen to coincide with the true surface mass density at this point. The dashed curves are also reconstructed mass profiles, which are related to the solid curve by the invariance transformation (3.29) (for $\lambda = -0.3, 0.5, 0.9, 1.1$). The outermost point shown corresponds to $4\rlap{.}'5$

for the reconstruction in the non-critical case. The cluster parameters are indicated in the caption, and, as always in this paper, we have chosen $D_{\mathrm{ds}}/D_{\mathrm{s}} = 0.6$. The upper panel compares the theoretical distortion function $\delta(x)$ with the one obtained by the preceding smoothing procedure. The amplitude of the variations depend of course on the smoothing scale. In the lower panel, we have plotted the true mass profile (dotted curve), and the profile obtained by applying (5.3) and (5.4) to the distortion $\delta$ as plotted in the upper panel (solid curve). The value of $\kappa(x_{\mathrm{out}})$ was chosen such that it agrees with the actual value at this point. It can be clearly seen that the relative variations of the reconstructed profile are much smaller than the corresponding variations in $\delta$, i.e., the reconstruction smoothes out noise in the distortion. The dashed curves in Fig. 10b also show mass profiles which yield the same distortion profile $\delta(x)$ as the solid curve, and are related to the latter by the invariance transformation (3.29), with four different values of $\lambda$. It can be seen that both 'reasonable' and 'unreasonable' (negative surface mass density – for $\lambda \gtrsim 1/(1-\kappa_{\rm out})$, or profiles which increase outwards – for $\lambda < 0$) mass profiles are compatible with the distortion $\delta(x)$. Hence, without further 'external' constraints, one cannot distinguish between the various profiles shown in Fig. 10b from the distortion data. The mass inside a radius $x$ behaves like

$$m(x,\lambda) = \lambda m_0(x) + (1-\lambda)x^2 \qquad (5.5)$$

under the transformation (3.29). Besides this basic ambiguity, which is inherent in all cluster reconstructions, the reconstructed profile follows the true profile remarkably well. Due to the small number of galaxy images close to the center of the cluster, the innermost part of the mass profile in not very well constrained; we shall see this to be the case also for the critical clusters to be discussed next.

When the cluster is critical, three essential problems occur. The first is the correct choice of the sign in (5.4), the second is the singular behaviour of (5.2) near $g=1$, and the third is related to the fact that near the radial critical curve, there are at most a few galaxy images, so that the profile $\delta(x)$ near the radial critical curve (and, more generally, close to the center) is not well determined from observations. We shall deal with these problems in turn. For illustration, we consider, as before, a cluster of the form (4.7) with velocity dispersion $\sigma_v = 1200\,{\rm km/s}$, but this time with a core radius of $0\rlap{.}'2$, which implies a central surface mass density of $\kappa_0 = 2.07$. Since the outer part of the cluster is nearly the same as that of Fig. 10, and the reconstruction of the outer parts of the cluster is unaffected by the the mass profile in the inner part, we choose an outer radius of $2\rlap{.}'5$ in which we distribute galaxies with a surface mass density of $40/({\rm arcmin})^2$. In Fig. 11a, we have plotted the true distortion $\delta$ as dashed curve, and the reconstructed distortion as obtained from smoothing as the solid curve. It can be seen that the smoothed version follows the true curve very closely, with the largest discrepancy occurring close to the radial critical curve. This has two reasons: first, the number of galaxy images within the radial critical curve is small (in the realization shown in Fig. 11, there are only 10 galaxy images with $x \leq 0.5$), and second, the distortion $\delta$ shows a sharp minimum at $x = x_{\rm r}$, which will be very much softened by a smoothing procedure.

Taking the smoothed profile as it is, we can calculate $g_i$ from (5.4). In a first trial, we use the minus-sign in (5.4) throughout, which is equivalent of pretending to have no critical curve. Fig. 11b shows the corresponding reconstructed mass profile in this case. Outside the tangential critical curve, the fit is as good as the one shown in Fig. 10, but it becomes grossly wrong for $x \leq x_{\rm t}$, as expected, when compared to the true mass profile, which is shown as dotted curve. Since the inversion relation (5.2) is singular near $g = 1$, and since the reconstructed distortion profile $\delta(x)$ is very close to unity near the tangential critical curve, a sharp rise occurs in the reconstructed profile at $x = x_{\rm t}$. Interestingly, the profile reaches a value close to, but smaller than, unity at the center, and *increases* outwards near the center. This is due to the fact that $\delta < 0$ near the center. From (4.4) we see that $\delta < 0$ can occur either if $\kappa > 1$ or $\Gamma < 0$. Since $\Gamma = \bar\kappa - \kappa$, $\Gamma < 0$ implies $\bar\kappa < \kappa$, which can only occur if the $\kappa$ increases with radius.

It should be noted that the density profile shown in Fig. 11b provides a perfect fit to the smoothed distortion function and, taken at face value, is an acceptable reconstruction.

The reason why one would reject this reconstruction is on physical grounds: the projected density of a cluster is not expected to rise with radius. Also, the sharp rise in $\kappa$ at $x = x_\mathrm{t}$ would appear artifical. Since the reconstructed distortion $\delta$ comes very close to unity, one could at least suspect that the cluster has a critical curve.

Thus, we next assume that there is a tangential critical curve near the point where the reconstructed distortion profile has its maximum. In this case, we shall try to locate the critical curve by fitting a distortion $\delta(x)$ near the maximum to the original data. This is done by noting that at $x = x_\mathrm{t}$, the distortion must locally behave as

$$\delta_\mathrm{fit} = 1 - \frac{2}{x_\mathrm{t}^2}(x - x_\mathrm{t})^2 + \mathcal{O}\left((x - x_\mathrm{t})^3\right) \quad . \tag{5.6}$$

This behaviour can be obtained from (5.2) by expanding $\kappa$ and $g$ around $x_\mathrm{t}$ and noting that $m(x_\mathrm{t}) = 1$; this then yields that $\mathrm{d}g/\mathrm{d}x = -2/x_\mathrm{t}$ at $x = x_\mathrm{t}$. Inserting this result into (3.4) then yields (5.6). We searched the location of the critical curve in the following way: for each trial value of $x_\mathrm{t}$ we have taken the three nearest galaxy images on both sides and fitted the distortion (5.6) to these images. The value of $x_\mathrm{t}$ chosen corresponds to the minimum of the error

$$E = \sum_{i=1}^{6} \frac{[\chi_{\mathrm{t}i} - \delta_\mathrm{fit}(x_i)]}{\sigma_\chi^2\left(\delta_\mathrm{fit}(x_i)\right)} \quad , \tag{5.7}$$

where the sum extends of the six nearest galaxy images, $\chi_{\mathrm{t}i}$ are their tangential alignments, and $\sigma_\chi(\delta)$ is defined in (3.17) and plotted in Fig. 2; for the fit, we used the approximation $\bar{\sigma}_\chi$, as given in (3.23). With this procedure, the location of the tangential critical curve can be accurately determined. Due to the small number of galaxy images near the radial critical curve, a similar procedure near $x_\mathrm{r}$ is not practical. We estimated $x_\mathrm{r}$ simply as the minimum of the smoothed distortion $\delta$.

Having determined $x_\mathrm{t}$ and $x_\mathrm{r}$, we can now determine the sign to be used in (5.4), i.e., we take the minus-sign for $x > x_\mathrm{t}$ and $x < x_\mathrm{r}$, and the plus-sign for $x_\mathrm{r} < x < x_\mathrm{t}$. With this choice of $g_i$, we have applied the recursion relations (5.3) and obtained the radial mass profile shown in Fig. 11c. One can see that this profile is even worse than that shown in Fig. 11b, and the reason for this can be easily understood. At the critical curves, the sign in (5.4) changes, which implies that $g$ changes from values $|g| < 1$ to $|g| > 1$. Since $g$ is a very sensitive function of $\delta$ for $|\delta|$ close to unity, this change is related to a large change in $g$ across the critical curves, even if $|\delta|$ comes very close to unity, at least near $x_\mathrm{t}$. To illustrate this quantitatively for the case shown in Fig. 11, the maximum of the smoothed distortion is $\delta = 0.98885$, so that $g$ changes from 0.86 to 1.16 across the tangential critical curve. Even more drastic is the jump across the radial critical curve from $-0.53$ to $-1.90$, with the minimum of $\delta$ being $-0.825$. These jumps of $g$ cause the discontinuities of the reconstructed profile in Fig. 11c. Combined with the singular behaviour of (5.2) near $x_\mathrm{t}$, the resulting profile even leads to negative values of $\kappa$ near the center of the cluster (not shown in the figure). But again we note that the profile in Fig. 11c provides a perfect fit to the smoothed distortion $\delta(x)$, so that this profile must again be rejected on *physical* grounds. Here, the negative values of $\kappa$ are unacceptable, as well as the discontinuities in $\kappa$. Additionally, plotting $g(x)$, the discontinuities are very obvious, and sufficient to reject the final result. In fact, for every given distortion function $\delta(x)$, one can obtain an infinite number of mass profiles which provide exact solutions of (5.2); two of them are shown in Figs. 11b,c. To generate more, one can just arbitrarily choose a change of sign in (5.4) at any $x$; the resulting profiles are extremely badly behaved, in that they have discontinuities and negative surface mass density, but they are perfect fits to the original distortion function.

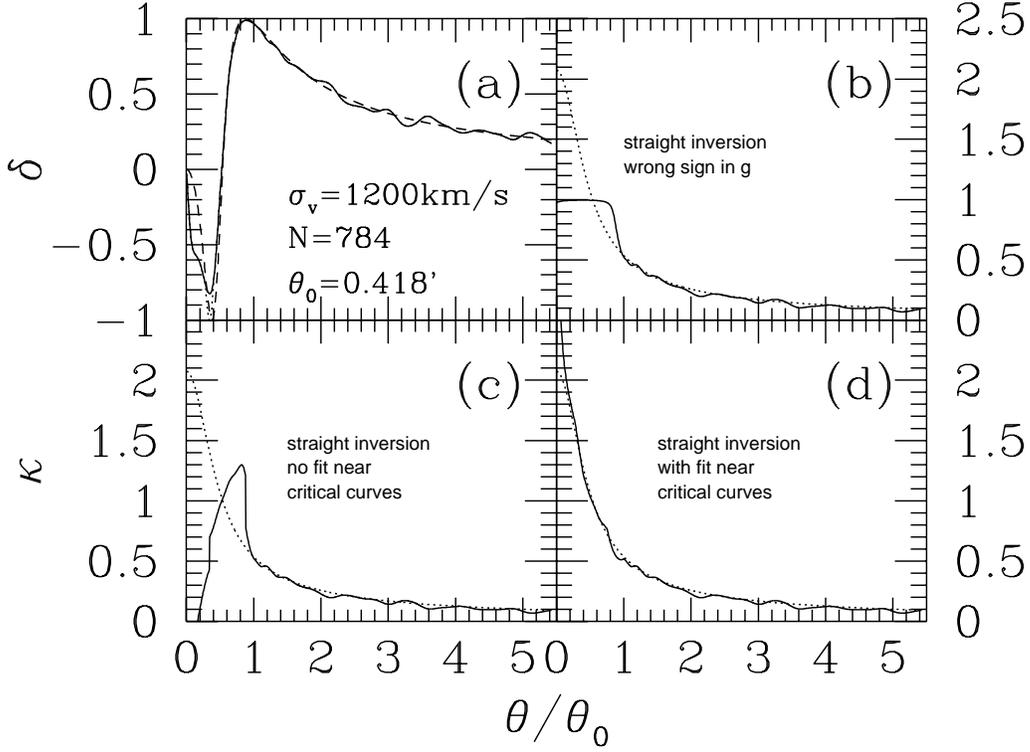

**Fig. 11.** For the same cluster model as in Fig. 10, except that now the core radius is $0.\!'2$, galaxies with the density of $40/\text{arcmin}^2$ and the internal ellipticity distribution (3.21) with $R = 0.3$ were mapped within in the inner $2.\!'5$. In the realization shown, this leads to a total of $N = 784$ galaxy images. The cluster is critical. (a) The true distortion $\delta(x)$ (dashed curve), the smoothed reconstructed distortion (solid curve), and the fitted distortion (dotted curve), which deviates from the smoothed distortion near the critical curves. (b) Solid curve shows the reconstructed density profile $\kappa(x)$, obtained from the smoothed distortion by assuming a negative sign in (5.4) for all $x$. (c) Reconstructed density profile from the smoothed distortion, with the correct sign in (5.4). (d) Reconstructed density profile from the smoothed distortion, but with a fit procedure near the critical curves (for more details, see text). In panels (b)–(d), the dotted curve is the true density profile. In all reconstructions, the outermost point was fixed to agree with the true density, to fix the transformation freedom (3.29)

In order to deal with the difficulties encountered near the critical curve, we have used a fitting procedure for the distortion near $x_\text{t}$ and $x_\text{r}$. We proceed in the following way: let $x_1$ and $x_2$ be two radial coordinates with $x_1 < x_\text{t} < x_2$; in the interval $x_1 \leq x \leq x_2$, we expand the mass $m(x)$ as a Taylor series,

$$m(x) \approx m_\text{fit}(x) = x_\text{t}^2 + \sum_{i=1}^{3} a_i \, (x - x_\text{t})^i \quad . \tag{5.8}$$

This ansatz automatically satisfies (5.7). The three coefficients $a_i$ are determined by the conditions

$$m_{\text{fit}}(x_2) = m(x_2) \quad ,$$

$$\frac{\mathrm{d}m_{\text{fit}}(x_2)}{\mathrm{d}x} = 2x_2 \kappa(x_2) \quad , \tag{5.9}$$

$$\frac{\mathrm{d}m_{\text{fit}}(x_1)}{\mathrm{d}x} \frac{1 - g(x_1)}{2x_1} = \frac{m_{\text{fit}}(x_1)}{x_1^2} - g(x_1) \quad ,$$

where the third condition follows from (5.2) at $x_1$, and $m(x_2)$ and $\kappa(x_2)$ are determined from the inward iteration (5.3). Similarly, let $x_3 < x_{\text{r}} < x_4$ be two coordinates on either side of the radial critical curve; then, the surface mass density for $x_3 \leq x \leq x_4$ is fitted by

$$\kappa(x) \approx \kappa_{\text{fit}}(x) = \sum_{i=0}^{3} b_i \, x^{2i} \quad , \tag{5.10}$$

and the four coefficients $b_i$ are obtained from the conditions

$$1 + \bar{\kappa}_{\text{fit}}(x_{\text{r}}) - 2\kappa_{\text{fit}}(x_{\text{r}}) = 0 \quad ,$$

$$\kappa_{\text{fit}}(x_4) = \kappa(x_4) \quad ,$$

$$m_{\text{fit}}(x_4) = m(x_4) \quad , \tag{5.11}$$

$$\kappa_{\text{fit}}(x_3)\bigl[1 - g(x_3)\bigr] = \frac{m_{\text{fit}}(x_3)}{x_3^2} - g(x_3) \quad ,$$

where $m_{\text{fit}}$ and $\bar{\kappa}_{\text{fit}}$ are obtained from $\kappa_{\text{fit}}$ via Eqs.(4.1 & 3). The conditions (5.9) and (5.11) lead to algebraic equations for the coefficients, which can be inverted analytically. Thus, within the intervals $x_1 \leq x \leq x_2$ and $x_3 \leq x \leq x_4$, we use the approximation formulae (5.8 & 10) instead of (5.3). The dotted curve in Fig. 11a shows the corresponding equivalent distortion $\delta$ in these intervals, whereas Fig. 11d displays the reconstructed density profile, using this fitting method. Except very close to the center, where the distortion information is sparse, we now obtain a very satisfactory fit for all radii. The result now depends on the values of $x_j$, $1 \leq j \leq 4$ for the fitting intervals; they have been chosen 'by eye' to get a fairly smooth connection of the fitted distortion to the smoothed distortion curve. If the interval is badly chosen, the reconstructed density profile will again show nearly discontinuous behaviour.

Whereas one can probably invent fitting formulae near the critical curves which are more sophisticated that (5.8 & 10), it is probably not worth the effort: the more advanced the fitting, and the larger the fitting intervals, the more the method deviates from a true direct inversion technique. Since the function $\delta$ has sharp features, it is much easier to obtain fits for $\kappa$ directly than fitting $\delta$ and determine $\kappa$ from this fit. Thus, we now turn to fitting methods.

### 5.2 Fitting by piecewise linear profiles

A simple way to fit a mass profile to the galaxy ellipticities is the following: select $n$ points $x_j$ and assume that the profile can be described, to sufficient accuracy, by a piecewise linear function, between $x_j$ and $x_{j+1}$. Then, the mass profile is completely determined by the values $\kappa(x_j)$, $1 \leq j \leq n$. In order to determine the $\kappa(x_j)$, one can minimize the error measure

$$\chi^2 = \sum_{i=1}^{N} \left( \frac{\delta(x_i) - \zeta(\delta(x_i))\chi_{\text{t}i}}{\sigma_\chi(\delta(x_i))} \right)^2 \tag{5.12}$$

with respect to the $\kappa(x_j)$, where the function $\zeta(\delta)$ and $\sigma_\chi(\delta)$ have been discussed in Sect. 3, and $\delta(x_i)$ is the value of the distortion calculated from the piecewise linear mass profile. One way to achieve this minimization is by the Levenberg–Marquardt method, as described in Press et al. (1992); we used the implementation `mrqmin` as described in this book. The only difference to the standard $\chi^2$ minimization is that the dispersion $\sigma_\chi$ is described by the model, and not by the data. So the $\sigma_\chi$'s have to be calculated from the current mass model at each step of the iteration.

For simplicity, we have chosen an equidistant grid in $x$, with $x_j = (j-1)x_{\rm out}/n$, and have $\kappa(x_{\rm out})$ chosen to coincide with the true value, in order to fix the freedom of transformation (3.29). The quality of the fit can depend sensitively on the initial guess for the profile; we have taken as a starting model the $\kappa(x_j)$ from the profile shown as solid curve in Fig. 11d, to fit the same galaxy data. Within $x_{\rm out}$, there are $N = 632$ galaxy images (note that $x_{\rm out}$ corresponds to $0.9 \times 2\rlap{.}'5$), so a good fit should have a $\chi^2$ of about $N/2 - n$ (the factor $(1/2)$ is due to the fact that $\sigma_\chi$ was calculated for both components of $\delta$, whereas in the axially-symmetric case, only one component is fitted).

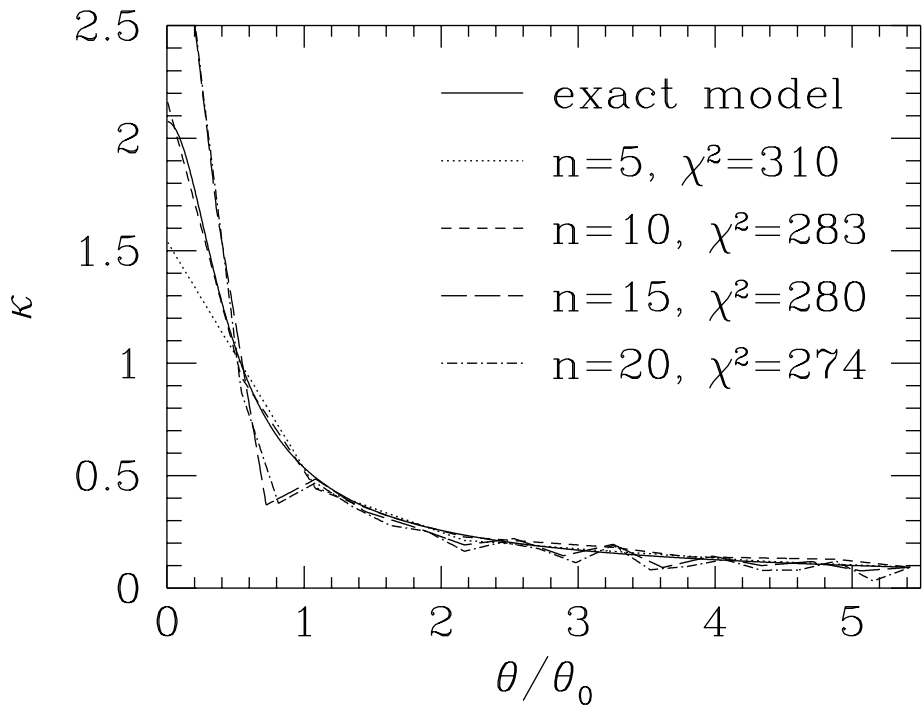

**Fig. 12.** For the same cluster and galaxy data as that used for Fig. 11, piecewise linear models for $\kappa$ are shown, for several values of the number $n$ of grid points

We have used this procedure to obtain profile fits for the same data on which Fig. 11 is based, for the values $4 \leq n \leq 20$. For each $n$, we have calculated the resulting value of $\chi^2$, which can change drastically with $n$, not necessarily monotonically. If we insert the distortion of the underlying mass profile in (5.12), we obtain a value of $\chi^2 \approx 289$. For four different values of $n$, the profiles are plotted in Fig. 12. Both, from the figure and the corresponding values of $\chi^2$, we see that the fit improves as we go from $n = 5$ to $n = 10$, but does not improve very much for larger $n$. In fact, choosing larger values of $n$ makes the profile to follow the fluctuations in the values of $\chi_t$ of the galaxy images, as seen by the fluctuating profiles for $n = 15, 20$. In fact, even if the underlying profile were not known, one would still accept the $n = 10$ fit to be better than the others shown. All curves cross the underlying profile at around $\kappa = 1$, so this point provides a strong constraint on the fit.

### 5.3 Fitting by other functions

The procedure described in the foregoing subsection can be generalized by fitting parametrized profiles to the ellipticity data. The piecewise linear profile has the advantage that no prejudice enters the fit, whereas a parametrized mass profile always implicitly makes assumptions about the true profile. On the other hand, the error measure (5.12) can control whether the best-fit model provides a sufficiently good description for the ellipticity data.

We choose here a simple example to show how the method works. Consider the profile fit

$$\kappa(x) = \sum_{j=1}^{n} \frac{a_n}{(b_n^2 + x^2)^{\alpha_n}} \quad ; \tag{5.13}$$

this fit formula contains $3n$ free parameters. We have fitted the data used for Figs. 11 and 12 with (5.13), with $n = 3$, again by minimizing (5.12) with the help of the routine `mrqmin` in Press et al. (1992). The result is shown in Fig. 13. Note that in this case, the parameter $\lambda$ in the invariance transformation (3.29) has not been fixed; instead, the smooth power-law behaviour of the fit formula (5.13) for $x \gg b_j$ effectively fits a (sum of) power-laws to the data at large radius.

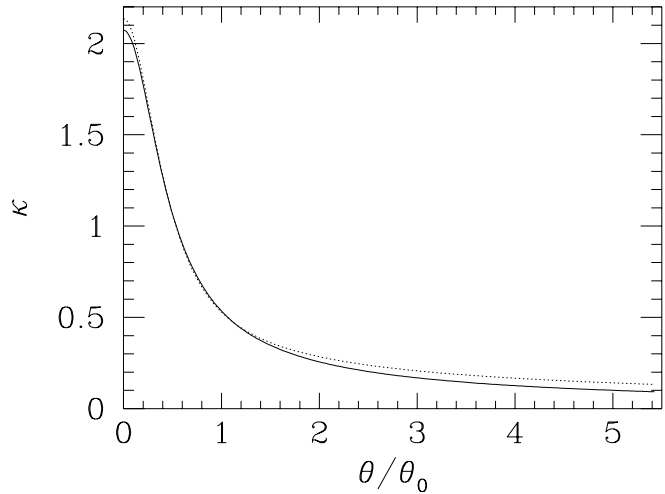

**Fig. 13.** For the same cluster and galaxy data as that used for Figs. 11 and 12, the parametrized profile (5.13) with $n = 3$ was fitted to the ellipticity data. The resulting fit is shown as dotted curve, whereas the underlying mass profile is shown as solid curve. The value of $\chi^2$ is 285, whereas if the true distortion $\delta$ is used in (5.12), the corresponding value is 289. The resulting fit parameters are $a_i = (0.441, 0.053, 0.150)$, $b_i = (0.411, 0.297, 0.679)$, $\alpha_i = (0.361, 0.850, 2.282)$

As seen from Fig. 13, the reconstructed profile matches the underlying mass distribution very well. Although the latter one can be represented exactly with the fit formula (5.13) for $n = 2$ ($a_1 = a_2 = 1/2$, $b_1 = b_2 = x_c = 0.482$, $\alpha_1 = 1/2$, $\alpha_2 = 3/2$), the minimization yields quite different values of these parameters (although the final result will depend on the initial choice of model parameters). The fit at larger radii can be improved by taking data in a larger field.

### 5.4 Discussion

The three different methods discussed in this section have their strengths and weaknesses. Certainly the strength of the direct inversion is that it makes no assumption about the underlying mass profile, although, as pointed out several times, there remains an arbitrariness with respect to (1) choosing the parameter $\lambda$ in the invariance transformation (3.29) and (2) the local choice of the sign in (5.4) due to the degeneracy between $g$ and $1/g$. An additional advantage of the direct method is that the intrinsic ellipticity distribution of the sources need not be known, whereas for the fitting method, the function $\sigma_\chi(\delta)$ must be known, which depends on $p_s(\chi^{(s)})$. However, for critical clusters, the direct inversion method has problems near the critical curves, so we were forced to introduce a local fitting of $\delta$ (or $\kappa$) close to them in order to obtain the satisfactory fit shown in Fig. 11d. These problems do not occur in the fitting procedures dealt with in Sects. 5.2 & 5.3, though they are burdened with different problems. Certainly, the method of Sect. 5.3 uses a prejudice about the general shape of the surface mass density, which can however be justified a posteriori by considering the value of $\chi^2$ of the final result. The piecewise linear profiles treated in Sect. 5.2 do not introduce such a prejudice. Both fitting methods have the technical problem that the minimization procedure can locate local minima, where they get stuck, or where they run into a parameter regime where the Hessian matrix of $\chi^2$ with respect to the fitting parameters becomes singular. That means that the initial choice of parameters have to be tried out, and many different trials may be necessary before the solution converges to an acceptable fit.

Of course, combinations of these methods are possible. For example, the direct inversion proceeds from large $x$ to smaller ones; hence, it can be used without problems from $x_\mathrm{out}$ to a value of $x$ close to the first critical curve (and right to the center if the cluster is not critical), and the inner part can be fitted by a parametrized mass model. Whereas in the one-dimensional case considered here the direct inversion yields results comparable with model fitting, it remains to be seen which of these methods is the more successful one in a realistic two-dimensional case. We shall deal with that problem in a forthcoming paper of this series (Seitz & Schneider, in preparation).

## 6 Conclusions

Progress in cluster inversion is currently driven by improvements in the observational techniques; they have already achieved a level where significant information about the general mass profiles of clusters can be obtained (Fahlman et al. 1994; Smail et al. 1994a,b; Bonnet et al. 1994). Future observational progress, such as larger CCD arrays, larger telescopes with improved optics etc., can be expected and will make the cluster inversion from gravitational distortions the most reliable method to obtain information about the matter distribution in clusters of galaxies. Furthermore, the refurnished Hubble

Space Telescope will provide high-resolution images of distant faint galaxies, so that their intrinsic shape distribution can be expected to be determined, or at least to be very much constrained. With these observational prospects in mind, we want to initiate a program for the theoretical analysis of such improved data. The first step, cluster inversion in the linear regime, has been undertaken in the pioneering work by Kaiser & Squires (1993). In a forthcoming paper, we will show that even in the linear regime, the original Kaiser & Squires method can be substantially improved; in particular, the noise can be reduced.

The present paper should be considered as a second step, in which we build the ground for nonlinear inversion techniques. We have concentrated on the simplest possible case, namely assumed that all sources have the same redshift; this allows us to concentrate on the essential properties of the statistical estimate of the local distortion caused by the lens. Our main results can be summarized as follows:

(1) There is a global invariance transformation (3.29) of the surface mass density which leaves the observable distortion unchanged; this transformatin corresponds to adding a uniform sheet of matter and simultaneously scaling the density profile, and is similar to the 'magnification transformation' found by Gorenstein et al. (1988) in a different context.

(2) There is a local invariance transformation (3.3), which can also be written as $\gamma \to \gamma(1-\kappa)^2/|\gamma|^2$, where $\gamma$ is the (complex) shear.

(3) We have obtained the relation between the local distortion $\delta$ and the local average of the image ellipticities (3.16). Depending on the intrinsic shape distribution of the sources, the correction factor $\zeta(|\delta|)$ can deviate significantly from unity. We have provided simple fit formulae for this correction factor, as well as for the dispersion of $\delta$.

(4) The local distortion $\delta$ can also be determined from the orientation of the galaxy images only, without using the absolute value of their ellipticity. At least for small distortions, this method yields results which are nearly comparable to the one in which full information about the image shapes is used (see Fig. 5). Refinements of this technique should further reduce the corresponding error.

(5) Knowing the distribution of intrinsic source shapes allows the determination of the dispersion $\sigma_chi$ of the local shear. This fact has been used several times in the fitting procedures to obtain the mass profile of spherical clusters. On the other hand, the general shape of the function $\sigma_\chi(\delta)$ should not be very different from that shown in Fig. 2, even for a quite different distribution in intrinsic source shapes.

(6) We have investigated several methods to determine the center of an axially-symmetric matter distribution. For non-critical clusters, the accuracy with which the center can be determined is about 1/3 of the mean separation of the galaxy images; however, for critical clusters, the accuracy can be improved to a median value of less than one tenth of the mean image separation (see Fig. 8).

(7) We have obtained a fully nonlinear inversion of clusters with axially-symmetric surface mass density in Sect. 5. Whereas the nonlinear aspects have not produced any difficulty for clusters without critical curves, they cause substantial problems for the inner parts of critical clusters. The symmetric case considered here also provides a useful demonstration for the problems one encounters because of the local invariance transformation mentioned in (2) above. These problems also indicate the difficulties one has to anticipate if a truly two-dimensional matter distribution is to be reconstructed from the distortion field. An infinite number of matter distributions are compatible with a given set of galaxy ellipticities; these distributions differ by different local choices of the local invariance transformation (2) and the overall constant $\lambda$ in the global invariance transformation (3.29). Concerning the first of these, the favoured mass profile can be singled out only by physical or plausibility arguments (such as the requirement of continuity and smoothness of the density profile).

We expect that the local invariance (2) can be partially broken if a distribution of sources with redshift is considered. However, close to the critical curves where this invariance causes the largest problems, one probably cannot hope that an additional degree of freedom (source redshift) will ease the reconstruction of the mass profile. From image distortions alone, there is no way to break the global invariance transformation (3.29). However, the recently suggested method based on spectroscopy of the brighter galaxy images, to compare the joint distribution in redshift and magnitude of galaxies behind a cluster and in the field (Broadhurst, Taylor & Peacock 1994) can break the degeneracy related to this invariance; hence, this latter approach can possibly provide at least very useful supplementary information for cluster inversion.

Smail, I., Ellis, R.S., Fitchett, M.J. & Edge, A.C. 1994b, preprint.

Soucail, G., Fort, B., Mellier, Y. & Picat, J.P. 1987, A&A 172, L14.

Soucail, G., Mellier, Y., Fort, B., Mathez, G. & Cailloux, M. 1988, A&A 191, L19.

Tyson, J.A., Valdes, F. & Wenk, R.A. 1990, ApJ 349, L1.